\documentclass[aps,prd,showpacs,nofootinbib,floats,floatfix,preprintnumbers,groupedaddress,twocolumn]{revtex4-1}
\usepackage{graphicx,epsfig}
\usepackage{dcolumn}
\usepackage{bm}
\usepackage{latexsym}
\usepackage[table]{xcolor}
\usepackage{booktabs}
\usepackage{color}
\usepackage{tabularx}
\usepackage{ulem}
\usepackage{hyperref}
\usepackage{float}
\usepackage{tabularx}
\usepackage{color}
\usepackage{comment}
\usepackage{physics}
\usepackage{subfigure}
\usepackage{tikz}
\usepackage{hyperref}
\usepackage{colortbl}
\usepackage{listings}
\usepackage{amsmath,amsfonts,amssymb}
\usepackage{fancyhdr}
\usepackage{orcidlink}
\usepackage{natbib}
\usepackage{multirow}

\hypersetup{
    colorlinks   = true,
    urlcolor     = blue,
    linkcolor    = blue,
    citecolor    = red
}

\def \m{\mu}
\def \n{\nu}
\def \a{\alpha}
\def \b{\beta}
\def \g{\gamma}
\def \d{\delta}
\def \l{\lambda}
\def \e{\epsilon}

\def \s{\sqrt}
\def \be{\begin{equation}}
\def \ee{\end{equation}}
\def \ben{\begin{eqnarray}}
\def \een{\end{eqnarray}}
\def \o{\omega}
\def \nb{\nabla}

\def \p{\partial}

\def \L{\mathcal{L}_m}
\def \A{\mathcal{A}}
\def \r{\rho}
\def \G{\Gamma}
\def \non{\nonumber}
\def \T{\Theta}

\begin{document}

    \title{$f(Q,T)$ gravity: From early to late-time cosmic acceleration}

    \author{Surajit Das$^{1}$\orcidlink{0000-0003-2994-6951}}
    \author{Sanjay Mandal$^{2}$\orcidlink{0000-0003-2570-2335}}
    
    \thanks{Corresponding author, Email: sanjaymandal960@gmail.com}
    
    \affiliation{$^{1}$Department of Physics, Birla Institute of Technology and Science - Pilani, Hyderabad Campus, Hyderabad, Telangana-500078, India\\
    \\
    $^{2}$Faculty of Symbiotic Systems Science, Fukushima University, Fukushima 960-1296, Japan}

\begin{abstract}
\noindent
    \textbf{Abstract:} In this article, we explore the comprehensive narrative of cosmic evolution within a cosmological framework by utilizing a novel form of gravity known as generalized symmetric teleparallel gravity, denoted as $f(Q,T)$ gravity. Here, $Q$ represents the non-metricity scalar, while $T$ denotes the trace of the energy-momentum tensor. We present and analyze two distinct $f(Q,T)$ cosmological models, each characterized by its unique Lagrangian. Our investigation delves into the cosmological parameters of these models, scrutinizing various energy conditions, examining the inflationary dynamics of the early universe through scalar field formulations, and probing the mysterious nature of dark energy using statefinder diagnostics and $(\omega-\omega')$ phase space analysis. Ultimately, our findings offer a comprehensive account of cosmic evolution, spanning from the early universe to its late-time evolution.
\end{abstract}

\maketitle

\section{Introduction}
\noindent
    In light of significant advancements in the fields of astronomy, astrophysics, cosmology, data science, space science, and technology, recent observations have provided compelling evidence for the accelerated expansion of our universe, driven by an enigmatic entity referred to as `Dark Energy'\cite{Riess1,Perlmutter1,Perlmutter2,Spergel1,Komatsu}. The history of cosmic expansion reveals two distinct eras of accelerated expansion \cite{Capoz}. The early universe experienced an epoch of accelerated expansion known as inflation, and in the present cosmic epoch, the universe's expansion is primarily driven by the dominance of dark energy.

    The inflationary epoch and the late-time accelerated expansion can be explained by considering matter comprising a barotropic fluid with a pressure-energy density relationship expressed as $p=f(\rho)$, where $p$ and $\rho$ denote the pressure and energy density, respectively. This approach has been discussed in the context of barotropic fluids \cite{Nojiri1,Nojiri2}, and inflationary scenarios within the framework of viscous fluid models have also been explored \cite{Bamba}. Conversely, some researchers argue that current observational data do not provide a definitive prognosis for the universe's ultimate fate \cite{Astas}, prompting the exploration of various proposals to resolve issues related to dark energy \cite{Ratra,Caldwell1,Buchert,Picon,Tomita,Milton,Hunt,Easson,Radicella,Pavn,Pandey}, parallelly in the contexts of dark matter \cite{G1}, and black hole-wormhole geometries \cite{G2,G3,G4,G5,G6,G7}.

    One intriguing avenue to understand the nature of dark energy involves modifying the standard theory of general relativity, that portraying it as a geometrical property of the universe. These modified gravity theories are rooted in the modification of the Einstein-Hilbert action \cite{Palia}. Various intriguing modified theories of gravity exist, including $f(R)$ gravity \cite{Soti,fr}, $f(R,T)$ gravity \cite{Har}. Another notable modified theory of gravity is $f(T)$ gravity, also known as teleparallel equivalent general relativity (TEGR) \cite{Bengochea,Ferraro,Linder1,Cai}, where the action is an arbitrary function of the torsion scalar $T$, replacing the Ricci scalar $R$ in the action.

    Beyond the standard curvature and torsion representations in general relativity, symmetric teleparallel gravity introduces a novel perspective \cite{Nester}. In this formalism, a geometric variable is defined by a non-metricity scalar $Q$, giving rise to $f(Q)$ gravity \cite{Lavinia}. Extending the $f(Q)$ theory involves considering non-minimal coupling between gravitational interactions generated by the term $Q$ and the trace of the energy-momentum tensor $T$. This results in an action containing an arbitrary function of $f(Q,T)$. In the realm of $f(Q,T)$ gravity, extensive studies have been conducted, including investigations into various cosmological scenarios \cite{Nisha,Xu,Gad}, equation of state constraints \cite{Simran3}, cosmological perturbations \cite{Antonio}, inflationary models \cite{Maryam}, quintessence universes \cite{Koussour}, late-time cosmic acceleration \cite{Lalke}, energy conditions \cite{Simran1,Simran2}, geometries of wormholes \cite{Tayde}, holographic dark energy \cite{Asutosh}, and the reconstruction of the $\Lambda$CDM universe \cite{gaurav}.

    In this paper, we adopt a phenomenological approach to comprehensively explore the entire cosmic evolution of the universe. We introduce an ansatz for a well-motivated scale factor, allowing us to derive analytical solutions for energy density, pressure, equation of state parameters, and various pertinent physical consequences for different cosmological models within the framework of $f(Q,T)$ gravity. This analysis is conducted to elucidate both early and late-time cosmic acceleration within our chosen parametrization.

    This paper is organized as follows: Section \ref{II} provides a brief discussion on Symmetric Teleparallel Gravity (STEG) as a foundation. In Section \ref{III}, we briefly outline the mathematical formulation of $f(Q,T)$ gravity and present the gravitational field equations in a spatially flat FLRW-type spacetime. Section \ref{IV} addresses fundamental cosmic parameters, including the scale factor, Hubble parameter, and deceleration parameter, and proceeds to construct cosmological models within $f(Q,T)$ gravity for two distinct scenarios in Section \ref{V}. This section offers insights into the expressions and behaviours of energy density, pressure, and equation of state parameters. Section \ref{VI} explores energy conditions for the chosen models. Section \ref{VII} delves into the formulation of scalar field descriptions for our models. Sections \ref{VIII} and \ref{IX} introduce geometrical diagnostics aimed at distinguishing our model from other dark energy models. Finally, Section \ref{X} summarizes our findings and offers a brief discussion of their implications.

\section{Symmetric Teleparallel Equivalent General Relativity in a Nutshell}\label{II}
\noindent
    Since general relativity (GR) is based on the Riemannian manifold, it should be generalized to a more general geometric theory of gravity, which could explain the gravitational field as a more general geometric gravitational structure that should also be valid in the solar scale system. So, we may explain the various cosmological aspects of the universe from the early to the late time acceleration. In this generalization process, Weyl introduced the notion of non-metricity, a new geometric quantity in which the covariant derivative of the metric tensor is non-zero \cite{Weyl/1918}. In Weyl's theory, there is an extra connection called the length connection, which gives us an information on the length of a vector and doesn't contain any knowledge about the direction of a parallel transported vector field. After Weyl, more generalizations of gravity theories have been formulated [see for more in detail \cite{Di,Von,D,Z,M1,T,L,J,Z1,Z2}].

    Therefore, it is clear that GR can be presented at least in two equivalent formalisms. One is the well-known curvature representation, where the non-metricity and the torsion tensor vanish. Another one is called the teleparallel equivalent general relativity (TEGR), in which the curvature and the non-metricity vanish, but the torsion tensor does not. Fortunately, a relatively unexplored territory consists of another third equivalent representation of GR, namely symmetric teleparallel equivalent of general relativity (STEGR) or simply symmetric teleparallel gravity \cite{Nester}. In this theory, the curvature and the torsion vanish, but the non-metricity scalar ($Q$) is the basic geometrical variable that describes the properties of geometric gravitational interactions. It also describes the length-variation of a parallel transported vector. This formulation is completely geometric and covariant. As the curvature and torsion relate with affine connections, not dealing with spacetime, the covariant derivatives must commute due to the vanishing of curvature. Here in STEGR, the associated energy-momentum density is essentially the Einstein pseudotensor but becomes a true tensor in the geometric representation. STEGR was further developed into an arbitrary generic function of $Q$, i.e., $f(Q)$ gravity, which is also known as coincident general relativity (CGR) \cite{Lavinia}. CGR is described by the Einstein-Hilbert action, excluding the boundary term, which is underpinned by the spin-2 field theory. This particular construction also provides a starting
    point for the modified gravity theories and presents the early and late-times cosmological solutions of the universe. That's why this is a simpler geometrical structure to the affine connection which is fundamentally devastating gravity from any inertial character.

    The STEGR can also be represented by a general quadratic and parity-conserving Lagrangian with the help of Lagrangian undetermined multipliers for vanishing torsion and curvature \cite{Adak}. This particular Lagrangian is equivalent to the Einstein-Hilbert Lagrangian in standard GR for certain choices of the coupling coefficients. However, it was also shown that the field equations can be written as a system of Proca equations, which may be an interesting study for the propagation of gravitational-electromagnetic waves \cite{Mol}. Conroy et al, \cite{Conroy} studied the action, completely made up of the non-metricity tensor, and its contractions were decomposed into terms, which involved the metric and a gauge vector field. So, this paper \cite{Conroy} discussed the derivation of the exact propagator for the most general infinite-derivative, even-parity in the generally covariant STEGR theory of gravity. Also, the linear perturbations in flat space were analyzed in \cite{Conroy,Kovisto}, and \cite{Beltran}. The propagation of gravitational waves with their various properties like speed and polarization was studied in \cite{Soudi} for the various extensions of symmetric teleparallel gravity. For all the possible classifications of quadratic, first-order derivative terms of the non-metricity tensor in the framework of symmetric teleparallel gravity, Dialektopoulos et al,\cite{Dia} used the Noether symmetry approach. Basically, in \cite{Dia}, they used to reduce the dynamics of the system by choosing symmetries to find analytical solutions, and this model was invariant under point transformations in a cosmological background. 

    On the other hand, from the cosmological point of view, it was seen in the pieces of literature \cite{Lu,Laz} that the accelerating expansion of the universe is an intrinsic property of geometry, and we need not deal with any extra fields or exotic dark energy, under $f(Q)$ gravity consideration. Cosmology and the behaviour of cosmological perturbations in $f(Q)$ gravity were investigated in \cite{Lavi2}. Energy conditions, cosmography analysis, Buchdahl quark star formation, and wormhole solutions under the background of $f(Q)$ gravity can be found in \cite{Sanjay1}, \cite{Sanjay2}, \cite{Sneha}, and \cite{Mustafa1,Hassan1} respectively. For more in detail, one can see here \cite{Raja,De,Raja2,Sanjay3,Sanjay4}.

    By introducing in the framework of the metric-affine formalism, an extension of a new class of symmetric teleparallel gravity was considered in \cite{Harko}, where the geometry part, i.e., the non-metricity $Q$ is non-minimally coupled to the matter Lagrangian $L_{m}$. A Lagrangian of the form $L=f_1(Q)+f_2(Q)L_{m}$ leads to the non-conservation of the energy-momentum tensor and appearance as an extra force in the geodesicity, where $f_1$ and $f_2$ are two generic functions of $Q$, and $L_{m}$ is the considered matter Lagrangian. Several cosmological applications were considered for some specific functional forms of the functions like power-law and exponential dependencies of the non-minimal couplings, and the cosmological solutions lead to a late-times accelerating universe.

    Finally, one may consider another kind of extension of symmetric teleparallel gravity theory by considering an action that contains the non-minimal coupling between the geometry, i.e., $Q$, and the trace of the energy-momentum tensor $T$ instead of matter Lagrangian unlike in the previous one. By considering this kind of construction, one can easily describe the gravitational interactions in the presence of geometry-matter coupling, and the cosmological solutions can describe both the accelerating and decelerating evolutionary phases of the universe. So $f(Q,T)$ gravity can provide hopeful insights for the description of the early and late phases of our universe.

\section{Mathematical Formulation of $f(Q,T)$ Gravity}\label{III}
\noindent
    In the symmetric teleparallel $f(Q,T)$ gravity theory, by imposing the condition that the connection is symmetric so that for coincident gauge, the Levi-Civita connection $\hat{\G}^\l_{\m\n}=0$. So the disformation tensor $L^\l_{\m\n}$ can be written as \cite{epjc},
\be
    L^\l_{\m\n}=-\G^\l_{\m\n}\label{1},
\ee
    The non-metricity tensor is given by $Q_{\a\m\n}=\nb_\a g_{\m\n}$ and it can be also defined as,
\be
    Q=-g^{\m\n}\Big(L^\a_{\b\m}L^\b_{\n\a}-L^\a_{\b\a}L^\b_{\m\n}\Big)\label{2},
\ee
    where from Eq.\eqref{1}, the disformation tensor is written as,
\be
    L^\a_{\b\g}=-\frac{1}{2}g^{\a\l}\Big(-\nb_\l g_{\b\g}+\nb_\b g_{\g\l}+\nb_\g g_{\b\l}\Big).\label{3}
\ee
    The non-metricity conjugate or the superpotential of the model is given by,
\be
    P^\a_{\m\n}=-\frac{1}{2}L^\a_{\m\n}+\frac{1}{4}(Q^\a-\Bar{Q}^\a)g_{\m\n}-\frac{1}{4}\d^\a~_{(\m}Q_{\n)},\label{4}
\ee
    where we have used two known results of the trace of the non-metricity tensor i.e.,
\be
    Q_\a=g^{\m\n}Q_{\a\m\n}\equiv Q^{~\m~}_{\a~~\m}~;~~
    \bar{Q}_\b=g^{\m\n}Q_{\m\b\n}\equiv Q^{\m}_{~~\b\m}.\label{5}
\ee
    In addition, we define the non-metricity scalar $Q$ as,
\ben
    &&Q=-Q_{\a\m\n}P^{\a\m\n}\non\\
    &&=-\frac{1}{4}\Big(-Q^{\a\n\l} Q_{\a\n\l}+2 Q^{\a\n\l}Q_{\l\a\n}
    -2 Q^\l\bar{Q}_\l+Q^\l Q_\l\Big).\non\\
    \label{6}  
\een
    Now the corresponding action in $f(Q,T)$ gravity can be written as,
\be
    \A=\int \frac{1}{16\pi G}f(Q,T)\s{-g}~d^4x+\int \L\s{-g}~d^4x\label{7},
\ee
    where $f(Q,T)$ is a generic arbitrary function of $Q$ and the trace of the energy-momentum tensor is $T$, $G$ being the Newtonian universal gravitational constant. The matter Lagrangian can be considered as $\L$ and $g$ is the determinant of the metric tensor $g_{\m\n}$. Here the energy-momentum tensor is defined as,
\ben
    T_{\a\b}=-\frac{2}{\s{-g}}\dfrac{\d({\s{-g}\L})}{\d g^{\a\b}}.\label{8}
\een
    So by taking the variation of the action Eq.\eqref{7} w.r.t the metric tensor $g_{\m\n}$, we obtain the field equation of $f(Q,T)$ gravity as,
\ben
    -\frac{2}{\s{-g}}\nb_\l\Big(f_Q\s{-g}P^\l_{~~\m\n}\Big)-\frac{1}{2}f g_{\m\n}-f_Q\Big(P_{\m\l\b}~Q^{~~\l\b}_\n\non\\
    -2 Q^{\l\b}_{~~\m}P_{\l\b\n}\Big)=\big(8\pi G-f_T\big)T_{\m\n}-f_T\T_{\m\n},\non\\
    \label{9}
\een
    where $f_Q=\frac{df}{dQ}$ and $f_T=\frac{df}{dT}$. $\T_{\m\n}$ is the variation of energy-momentum tensor wrt the metric tensor, such that
\ben
    \T_{\m\n}\equiv g^{\a\b}\frac{\d T^{\a\b}}{\d g^{\m\n}}.\label{10}
\een
    It should be mentioned that the field equation, Eq.\eqref{9} is valid only for the coincident gauge \cite{Lavinia}.

    By assuming the universe is homogeneous and isotropic, we consider flat FLRW-type metric having the form:
\ben
    ds^2=-dt^2+a^2(t)\sum_i(dx^{i})^2.\label{11}
\een
    Here $a(t)$ describes the evolution history of the universe, simply known as scale factor and $i$ runs from $(1-3)$ only. Moreover, considering the cosmic fluid behaves as perfect fluid having the energy-momentum tensor as,
\ben
    T_{\m\n}=(\r+p)u_\m u_\n +pg_{\m\n},\label{12}
\een
    where $\r$, $p$, and $u^\m$ are the energy density, pressure, and the four-velocity of the cosmic fluid, respectively. The non-metricity $Q$ in this background can be calculated and it is given by $Q=6H^2$.

    Using the metric Eq.\eqref{11} in the field equations Eq.\eqref{9}, we can write the generalized Friedmann equations (taking $G=1$), which are
\ben
    8\pi \r=\frac{f(Q,T)}{2}-6FH^2-\frac{2\bar{G}}{1+\bar{G}}\big(\dot{F}H+F\dot{H}\big),\label{13}
\een
    and
\ben
    8\pi p=-\frac{f(Q,T)}{2}+6FH^2+2\big(\dot{F}H+F\dot{H}\big)\label{14}.
\een
    Here (·) dot represents a derivative with respect to cosmic time, $F=\frac{df}{dQ}$ and $8\pi\bar{G}=\frac{df}{dT}$.

    We can also write Einstein’s field equations from Friedmann equations as follows:
\be
    3H^2\equiv 8\pi \r_{eff}=\frac{f}{4F}-\frac{4\pi}{F}\big[(1+\bar{G})\r+\bar{G}p\big],\label{15}
\ee
\ben
    2\dot{H}+3H^2\equiv -8\pi p_{eff}=\frac{f}{4F}-\frac{2\dot{F}H}{F}\non\\
    +\frac{4\pi}{F}\big[(1+\bar{G})\r+(2+\bar{G})p\big]\label{16}.
\een
    Here $\r_{eff}$ and $p_{eff}$ are the effective energy density and effective pressure, respectively. These relations are helpful in exploring the effect of modified gravity as an alternative candidate to dark energy. As of now, we are looking forward to a complete scenario of the universe, including dark energy. Therefore, we will be working on total pressure and density.

    Using Eq.\eqref{13} and Eq.\eqref{14}, the equation of state parameter can be written as,
\ben
    \o=-1+\dfrac{4(\dot{F}H+\dot{H}F)}{f(1+\bar{G})-12FH^2(1+\bar{G})-4\bar{G}(\dot{F}H+\dot{H}F)}.\non\\
    \label{17}
\een

    Now, using the above setup, we'll explore various cosmological scenarios in the framework of $f(Q,T)$ gravity.

\section{Cosmic Parameters}\label{IV}
\noindent
    In this section, we shall discuss various cosmological parameters like the scale factor $a(t)$, the Hubble parameter $H(t)$ and the deceleration parameter $q(t)$. These particular kinematic variables play very crucial roles in the study of physical cosmology. Also, these parameters are the key parameters of various cosmological models in modified gravity theories, and these can describe the evolutionary history of the universe. 

    In this paper, we choose the scale factor $a(t)$ having the following form \cite{Moraes}: 
\begin{equation}
    a(t)=e^{ct}[\text{sech}^d(n-mt)].\label{18}
\end{equation}
    
    One important point is to be noted that this is not the only possible choice for the analysis. One can choose the scale factor according to their motivation and problem to work on. The main motivation for considering this specific form of the scale factor is that we are interested in exploring a complete profile of the universe, i.e., starting from early inflation to the late-time acceleration through the cosmological models. In addition, modified theories of gravity are well-known for being successful in describing dark energy as an alternative source; therefore, investigating all three profiles, such as inflation, matter-dominated, and late-time acceleration through a single factor in the context of modified theories, is interesting. Therefore, this is the only scale factor that allows us to explore and achieve this goal.

\begin{figure}[H]
  \centering
  \includegraphics[width=8.5 cm]{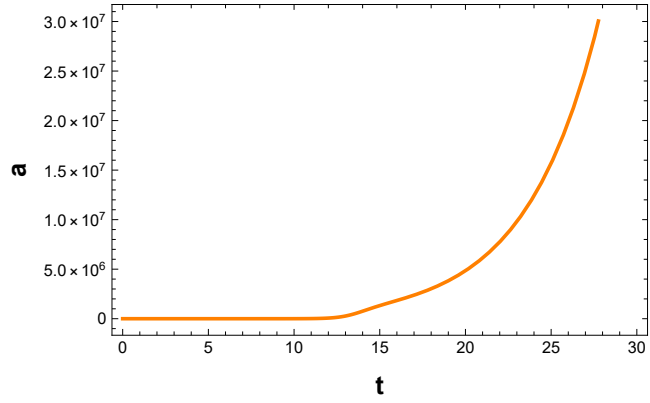}
  \caption{Plot of scale factor $(a)$ as a function of cosmic time $(t\, \text{in Gyr})$ for $c=0.97,d=1,m=0.735$ \& $n=10$.}
  \label{f1}
\end{figure}

    Using Eq.\eqref{18}, one can obtain the Hubble parameter $H(t)$ as,
\begin{equation}
    H(t)=\frac{\dot{a}}{a}=c+dm \tanh(n-mt).\label{19}
\end{equation}

\begin{figure}[H]
  \centering
  \includegraphics[width=8.5 cm]{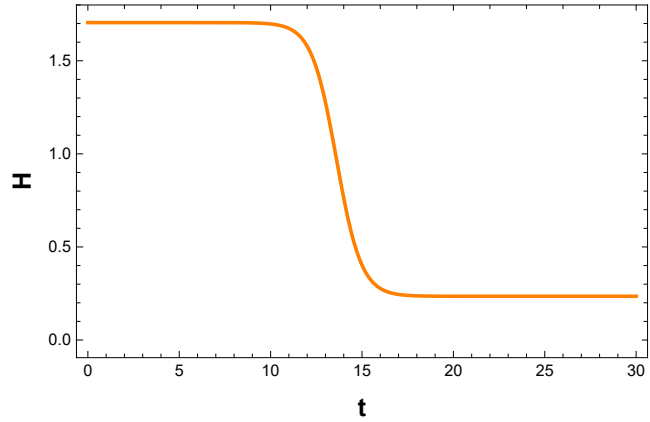}
  \caption{Plot of Hubble parameter $(H)$ as a function of cosmic time $(t\, \text{in Gyr})$ for $c=0.97,d=1,m=0.735,$ \& $n=10$.}
  \label{f2}
\end{figure}

    Also the deceleration parameter $q(t)$ is given by,
\ben
    &&q=-1-\frac{\dot{H}}{H^2}\non\\
    &&=-1+\frac{d m^2}{[c\cosh(n-mt)+dm\sinh(n-mt)]^2}.~~\label{20}
\een

\begin{figure}[H]
  \centering
  \includegraphics[width=8.5 cm]{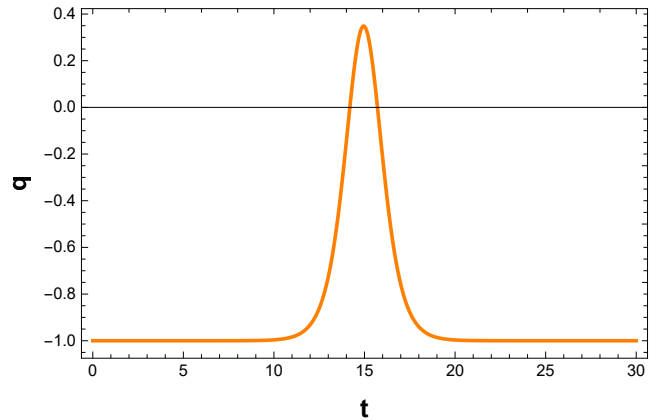}
  \caption{Plot of deceleration parameter $(q)$ as a function of cosmic time $(t\, \text{in Gyr})$ for $c=0.97,d=1,m=0.735,$ \& $n=10$.}
  \label{f3}
\end{figure}

    It is worth mentioning from the analysis of the research articles \cite{Riess1,Perlmutter1,Perlmutter2,Spergel1,Komatsu} that our universe is now accelerating. As an accelerated expansion phase, we know from the Friedmann equation that the second order derivative of the scale factor $\ddot{a}>0$ or $\dot{a}$ must be an increasing function over the time evolution. That means the Hubble parameter $H(t)$ is a decreasing function over the evolution of cosmic time. So Fig.\ref{f2} tells us that $H(t)$ was approximately the same in the early stage of the universe, and then it gradually decreased and finally kept a constant nature during the late time acceleration of the universe. The decreasing behaviour of the Hubble parameter with time may also escape the universe from phantom-like evolution. As the Hubble parameter is proportional to the energy density in the late times of evolution, here we get the same natural behaviour of the Hubble parameter in the late times of cosmic evolution.

    On the other hand, one can observe that the evolution of the deceleration parameter starts at $q=-1$, which represents the de-Sitter expansion phase, and then it goes to the deceleration phase through the accelerating power law expansion phase. After that, it again goes to the de-Sitter expansion phase in the late time of cosmic evolution (Fig.\ref{f3}). So $q(t)$ is positive during decelerating expansion and negative during accelerating expansion of the universe. Note that, for the eternally accelerated universe $q<0$, $q=0$ occurs in the transition line, and for the super-accelerated expansion, $q<-1$. This kind of behaviour of the universe is also suggested by the observed cosmological evolution of the universe \cite{Dodelson}.

\section{Cosmological Models in $f(Q,T)$ Gravity}\label{V}
\noindent
    In this present section, we discuss various functional forms of $f(Q,T)$ gravity model to analyze the cosmological evolutions in the symmetric teleparallel equivalent general relativity. From a mathematical point of view, one can see that the field equations are differential equations. Solving those equations, one can end up with either a power law, an exponential, or a combination of both $f(Q,T)$ forms. Therefore, without loss of generality and arbitrary choices, we choose two well-explored forms of $f(Q,T)$. The choices of $f(Q,T)$ models are as follows.

\subsection{Case~I~: $f(Q,T)=\a Q+\b T$.}
\begin{figure}[H]
  \centering
  \includegraphics[width=8.5 cm]{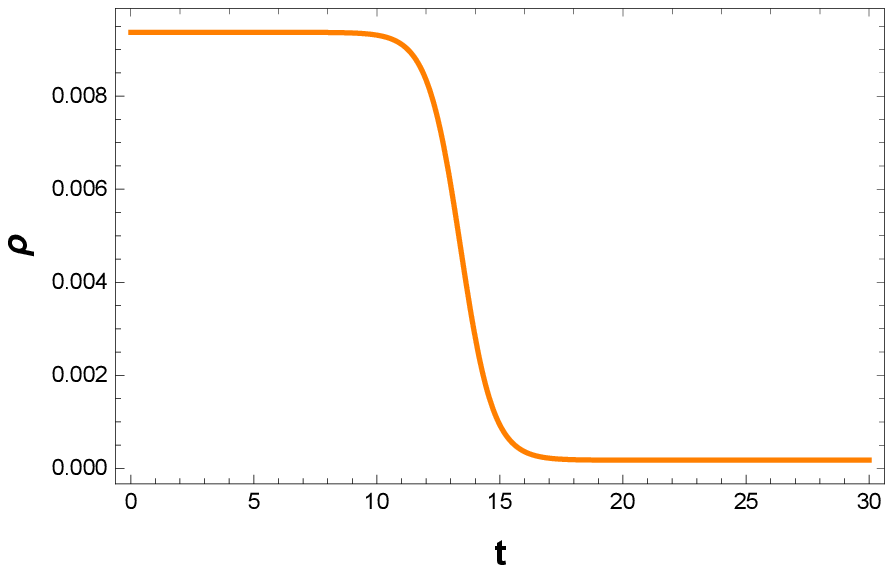}
  \caption{Plot of energy density $(\r)$ as a function of cosmic time $(t\, \text{in Gyr})$ for $c=0.97,d=1,m=0.735,$ \& $n=10$ with $\a=0.1$ and $\b=-59.1$.}
  \label{f4}
\end{figure}

    The first simple form of $f(Q,T)$ has 
\ben
    f(Q,T)=\a Q+\b T,\label{21}
\een
    where $\a$ and $\b$ are constants. So we obtain $F=F_Q=\a$, and $8\pi\bar{G}=f_T=\b$.
\\
    Now using the definition of scale factor mentioned in Eq.\eqref{18} and Eq.\eqref{21} in Eq.\eqref{13}, Eq.\eqref{14} and Eq.\eqref{17}, we have the following expressions for the energy density $\r$, pressure $p$ and equation-of-state (EOS) parameter $\o$. The plots of energy density $\r$, pressure $p$ and the equation-of-state parameter (EOS) $\o$ with the cosmic time $t$ is shown in Figs.\ref{f4}, \ref{5}, and \ref{6}, respectively.\\

\begin{widetext}
\ben
    &&\r=-\Big(\dfrac{1}{8\pi +2\b}\Big)\Big[3\a\big(c+dm\tanh(n-mt)\big)^2+\frac{\b d m^2\a}{8\pi +\b}~\text{sech}^2(n-mt)\Big],\label{22}
\\
    &&p=\dfrac{3\a\big[c+dm\tanh(n-mt)\big]^2}{\big(8\pi +2\b\big)}-\dfrac{\big(16\pi +3\b\big)~dm^2\a~\text{sech}^2(n-mt)}{\big(8\pi +2\b\big)\big(8\pi +\b\big)},\label{23}
\\
    &&\o=\dfrac{dm^2\big(16\pi+3\b\big)~\text{sech}^2(n-mt)-3\big(8\pi+\b\big)\big[c+dm\tanh(n-mt)\big]^2}{3\big(8\pi+\b\big)\big[c+dm\tanh(n-mt)\big]^2+\b dm^2~\text{sech}^2(n-mt)}.\label{24}
\een
\end{widetext}

\subsection{Case~II~: $f(Q,T)=u Q^{\e+1}+v T$.}
\noindent
    As another example of a cosmological model in the $f(Q,T)$ gravity, we consider a generic function given by,
\ben
    f(Q,T)=u Q^{\e+1}+v T,\label{25}
\een
    where $u,~v$ and $\e$ are constants. In this case, $F=f_Q=(\e+1)~u Q^\e=(\e+1)~u~6^\e H^{2\e}$ and $8\pi\bar{G}=v$. The generalised energy density $\r$, pressure $p$ and equation-of-state (EOS) parameter $\o$ in this case are given as the following:

\begin{widetext}
\ben
    &&\r=-\Big(\dfrac{1}{8\pi +2v}\Big)\big[c+dm\tanh(n-mt)\big]^{2\e}(1+2\e)u 6^\e\Bigg[3\big[c+dm\tanh(n-mt)\big]^2+\frac{v(1+\e)}{(8\pi +v)}~dm^2~\text{sech}^2(n-mt)\Bigg],\non\\
    \label{26}
\\
    &&p=\dfrac{3u~6^\e(1+2\e)\big[c+dm\tanh(n-mt)\big]^2}{\big(8\pi +2v\big)}-\dfrac{6^\e (16\pi+3v)(1+\e)(1+2\e)~u dm^2~\text{sech}^2(n-mt)}{\big(8\pi +2v\big)\big(8\pi +v\big)}\Big[c+dm\tanh(n-mt)\Big]^{2\e},\non\\
    \label{27}
\\    
    &&\o=-\dfrac{(8\pi+v)\big[c+dm\tanh(n-mt)\big]^{-2\e}3\big[c+dm\tanh(n-mt)\big]^2}{dm^2v(1+\e)~\text{sech}^2(n-mt)+3(8\pi+v)\big[c+dm\tanh(n-mt)\big]^2}\non\\
    &&+\dfrac{\big[c+dm\tanh(n-mt)\big]^{-2\e}dm^2(16\pi+3v)(1+\e)~\text{sech}^2(n-mt)\big[c+dm\tanh(n-mt)\big]^{2\e}}{dm^2v(1+\e)~\text{sech}^2(n-mt)+3(8\pi+v)\big[c+dm\tanh(n-mt)\big]^2}.\label{28}
\een
\end{widetext}    
    
\begin{figure}[H]
  \centering
  \includegraphics[width=8.5 cm]{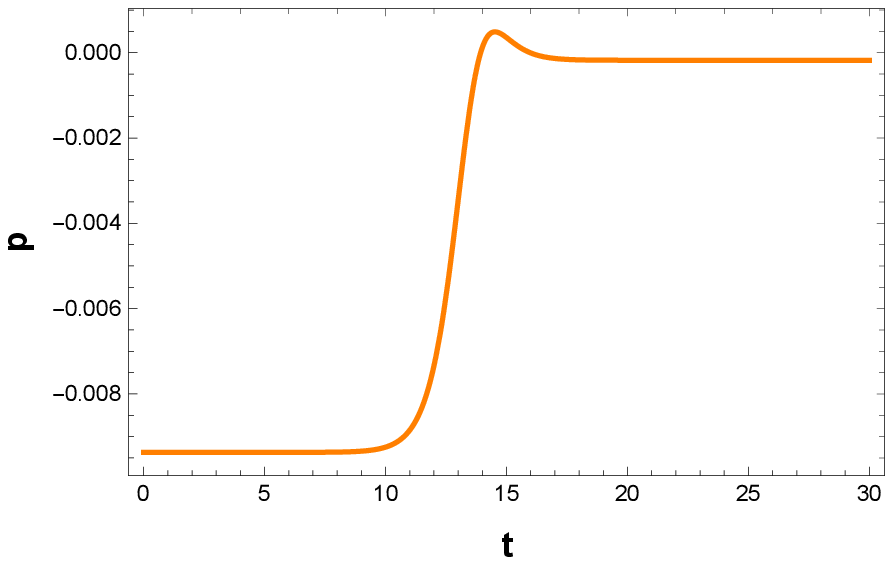}
  \caption{Plot of pressure $(p)$ as a function of cosmic time $(t\, \text{in Gyr})$ for $c=0.97,d=1,m=0.735,$ \& $n=10$ with $\a=0.1$ and $\b=-59.1$.}
  \label{f5}
\end{figure}

\begin{figure}[H]
  \centering
  \includegraphics[width=8.5 cm]{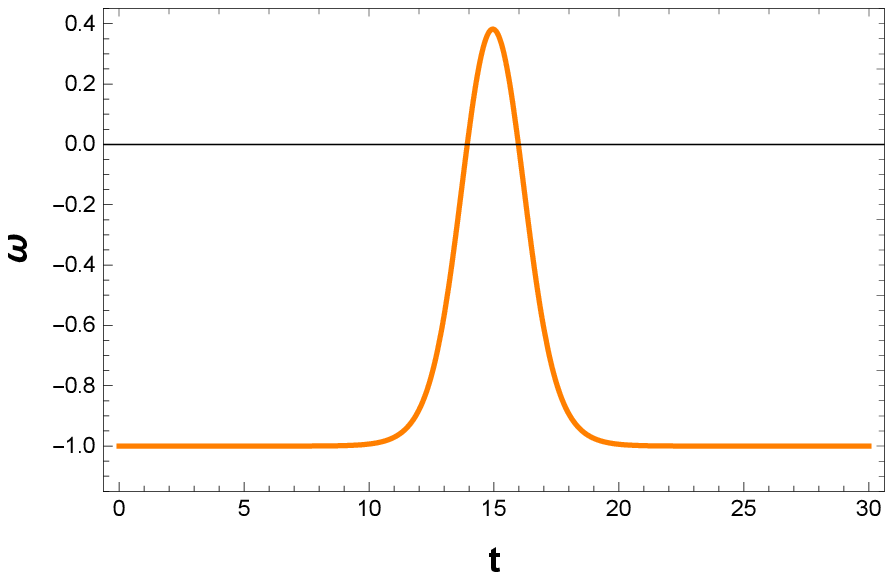}
  \caption{Plot of EOS parameter $(\o)$ as a function of cosmic time $(t\, \text{in Gyr})$ for $c=0.97,d=1,m=0.735,$ \& $n=10$ with $\a=0.1$ and $\b=-59.1$.}
  \label{f6}
\end{figure}
       
\begin{figure}[H]
  \centering
  \includegraphics[width=8.5 cm]{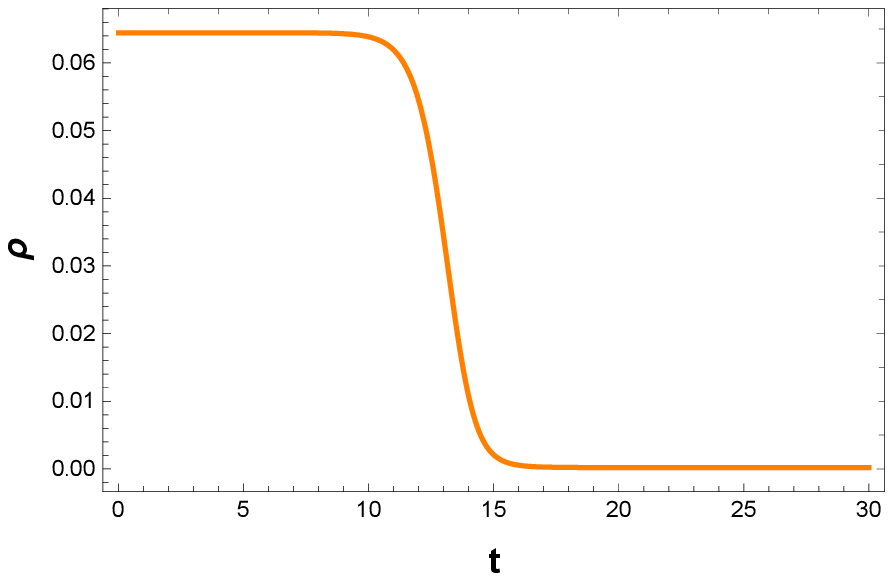}
  \caption{Plot of energy density $(\r)$ as a function of cosmic time $(t\, \text{in Gyr})$ for $c=0.97,d=1,m=0.735,$ \& $n=10$ with $u=0.1$ and $v=-59.1$.}
  \label{f7}
\end{figure}

\begin{figure}[H]
  \centering
  \includegraphics[width=8.5 cm]{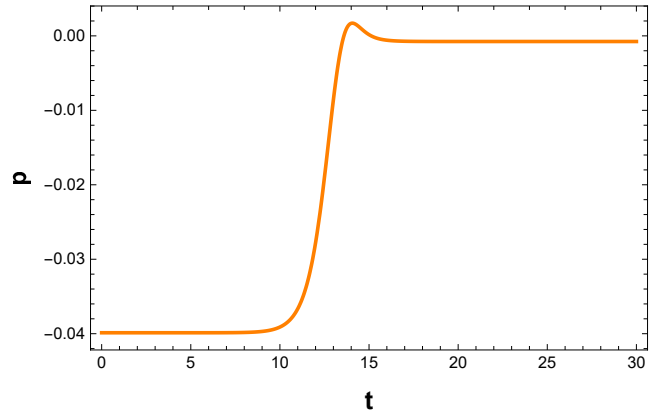}
  \caption{Plot of pressure $(p)$ as a function of cosmic time $(t\, \text{in Gyr})$ for $c=0.97,d=1,m=0.735,$ \& $n=10$ with $u=0.1$ and $v=-59.1$.}
  \label{f8}
\end{figure}

\begin{figure}[H]
  \centering
  \includegraphics[width=8.5 cm]{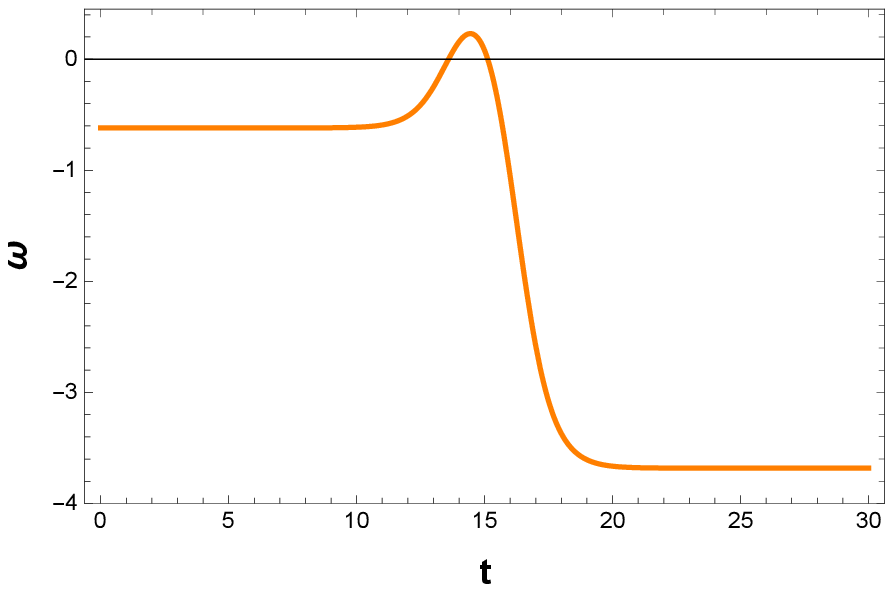}
  \caption{Plot of EOS parameter $(\o)$ as a function of cosmic time $(t\, \text{in Gyr})$ for $c=0.97,d=1,m=0.735,$ \& $n=10$ with $u=0.1$ and $v=-59.1$.}
  \label{f9}
\end{figure}

    It should be mentioned that for $\e=0$, the model given in Eq.\eqref{25} goes to the first example of $f(Q,T)$ model. All the corresponding energy density, pressure, and EOS parameter in Eq.\eqref{26}, Eq.\eqref{27}, and Eq.\eqref{28} reduce to the Eq.\eqref{22}, Eq.\eqref{23} and Eq.\eqref{24}, respectively. So the choice of $f(Q,T)=uQ^{\e+1}+vT$ behaves as a more general one. The profiles of energy density, pressure, and the EOS parameter vs. cosmic time with taking $\e=0.45$ is represented in Fig.\ref{f7}, \ref{f8} and \ref{f9}, respectively.

    Now, from the energy density profiles in the above two cases, it is observed that the energy density is high in the early times of the universe and then gradually decreases and is set to zero. On the other hand, the pressure $p$ lies in the negative range to suffice the acceleration of the universe, which could satisfy the dark energy dominated accelerating universe. For the behaviour of equation of state parameter $\o$, it is clear that $\o$ takes the value exactly $-1$ for the first case (Fig.\ref{f6}), whereas it takes value close to $-1(\approx -0.66)$ for the second example (Fig.\ref{f9}) in the very early time, after that it smoothly raises to its maximum value and then starts decreases again. From Fig.\ref{f6}, it is observed that $\o$ takes its maximum value, which is nearly close to $1/3$, and then it converges to exactly $-1$ to produce the recent observation \cite{Riess1,Perlmutter1,Perlmutter2,Spergel1,Komatsu}. The same case is happening for the general case in Fig.\ref{f9} under the consideration of the constant $\e=0.45$, but here $\o$ decreases to more negative values $\sim -3.8$ to serve the huge late-times acceleration, which could be generated by the negative pressure. Thus, we model a complete cosmic evolutionary history of the universe having inflation, radiation, matter, and dark energy dominated eras in order by using a simple ansatz choice of scale factor. Obviously, the ranges of the plot profiles of energy density, pressure, and the equation of state parameters for different cases are different. This motivates us to work with the different kinds of model consideration in $f(Q,T)$ gravity to study the cosmological scenarios. Finally, from Figs.\ref{f6} and \ref{f9}, it is also worth mentioning that the universe starts with the acceleration smoothly, goes to the deceleration phase, and finally returns to its late phase of accelerated expansion, produced by dark energy.

\section{Various Energy Conditions}\label{VI}
\noindent
    The energy conditions are the expressions of the linear relationships between the energy density $\r$ and pressure $p$. It has a very crucial significance in understanding the singularity theorem, describing the nature of geodesics, and studying the properties of black holes. The energy conditions are also the essential tools to study the geodesics of the Universe from the cosmological point of view. The physical conditions on the energy-momentum tensor are basically the energy conditions, which are obtained by making a connection between the Raychaudhuri equation \cite{Ray1,Ray2,Ray3} and the Einstein field equation.

    From the Raychaudhuri equation and by the constraints of gravity to be attractive in nature, we have the following two energy conditions, namely strong energy condition (SEC) and null energy condition (NEC) with a time-like vector field $v^\m$ and a null vector $k^\m$, respectively, i.e.,
\ben
    &&\big(T_{\m\n}-\frac{1}{2}T g_{\m\n}\big)v^\m v^\n\geq 0\implies \r+3p\geq 0~,~~~\textbf{SEC}\non\\
    \label{29}
\\
    && T_{\m\n}k^\m k^\n\geq 0\implies \r+p\geq 0,~~~\textbf{NEC}\label{30}
\een
    On the other hand, the weak energy condition (WEC) can be achieved by replacing null vector field $k^\a$ with a time-like vector $v^\a$ and the dominant energy condition (DEC) says that the matter should follow the time-like or null world lines.
\ben
    &&\r\geq 0~~with~~\r+p\geq 0~,~~~\textbf{WEC}\label{31}
\\
    &&\r\geq 0~~with~~\r\pm p\geq 0~,~~~\textbf{DEC}\label{32}
\een
    where $T_{\m\n}$ is the energy-momentum tensor and $g_{\m\n}$ is the fundamental metric tensor ; $T=T^{\m\n}g_{\m\n}$, being the trace. Various energy conditions in $f(Q,T)$ gravity have been studied in \cite{Simran1,Simran2}, but here we proceed to construct all the energy conditions on the background of Eq.\eqref{18}.

\subsection{Case-A : $f(Q,T)=\a Q+\b T$}
\noindent
    For the simplest choice of the $f(Q,T)=\a Q +\b T$ gravity model (where $\a,~\b$ are constants), we already have the expressions for the energy density $\r$ and pressure $p$ (Eq.\eqref{22}, and Eq.\eqref{23}). As we know, the energy density must be a positive quantity; here, we chose the model parameters $\a=0.1,~\b=-59.1$ to ensure the energy density is positive and the equation of state parameter as per the observations. Due to the importance of the energy conditions, we observed the acceptable ranges of model parameters $\a$ and $\b$, where the variation of $\a$ is from $9.1$ to $29.1$ and for $\b$, it is from $-25.55$ to $-55.55$.

\begin{widetext}
    \begin{figure}[H]
        \centering
	\begin{center} 
	$\begin{array}{cc}
	\subfigure[]
        {\includegraphics[width=1.0\linewidth,height=0.66\linewidth]{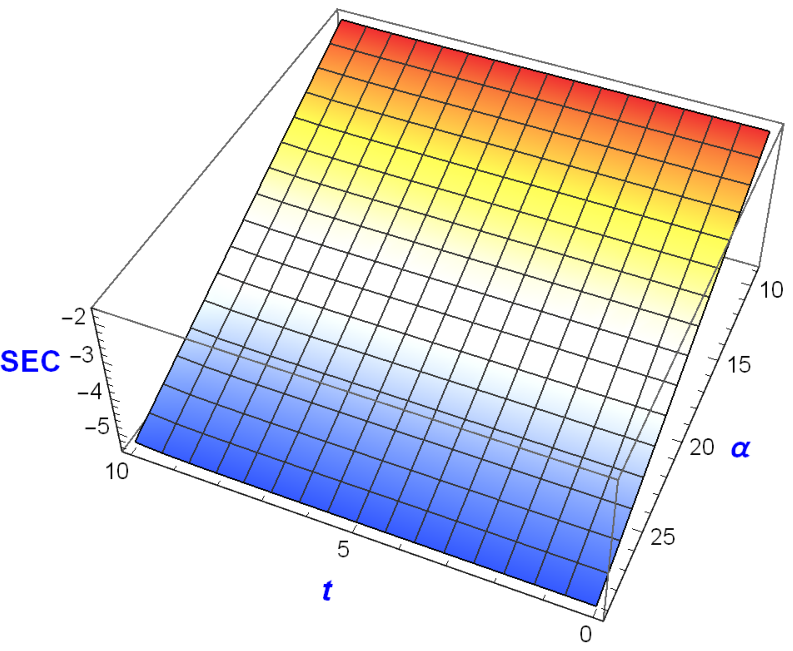}\label{10a}}
        \qquad
        \subfigure[]{\includegraphics[width=1.0\linewidth,height=0.66\linewidth]{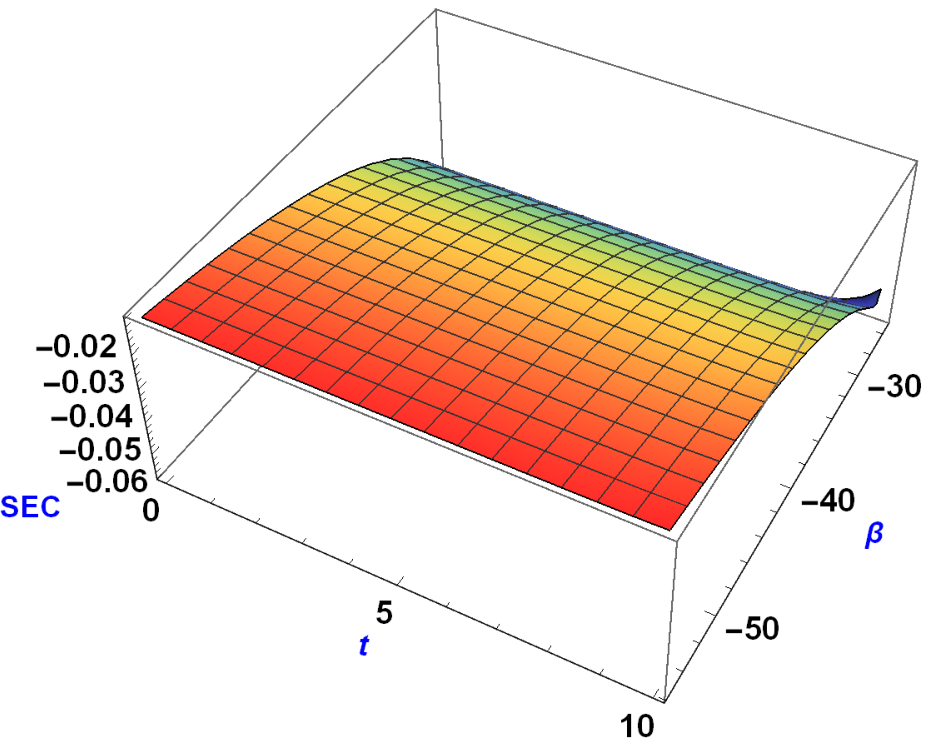}\label{10b}}
        \end{array}$
        \end{center}
        \begin{minipage}{\textwidth}
        \caption{Variations of SEC with $\a$, $\b$ and $t$.~$\text{Here $t$ is in Gyr.}$}\label{f10}
        \hrulefill
	\end{minipage}
        \begin{center}
        $\begin{array}{cc}
        \subfigure[] 
        {\includegraphics[width=1.0\linewidth,height=0.66\linewidth]{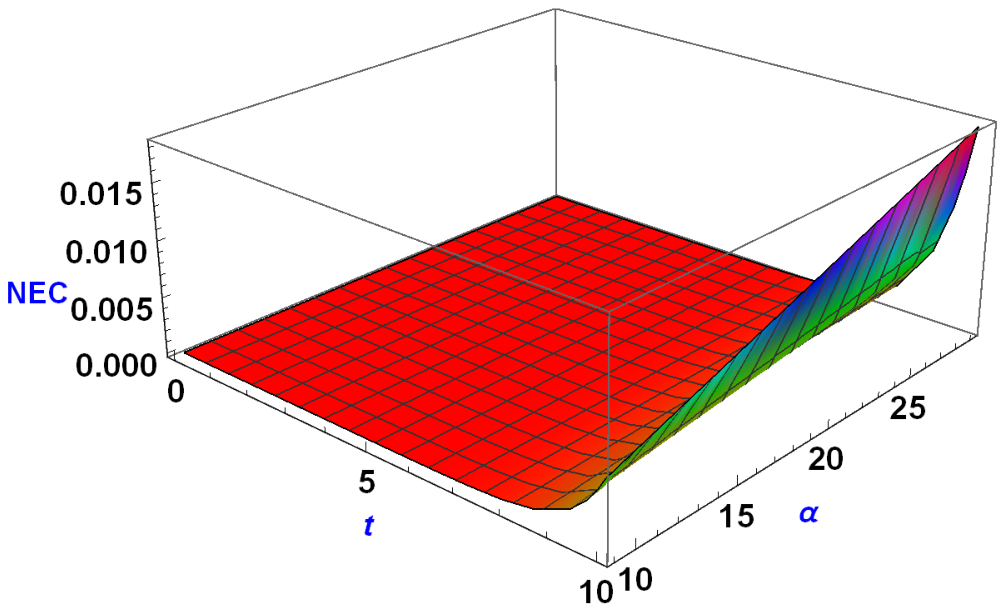}\label{11a}}
        \qquad
	\subfigure[] 
        {\includegraphics[width=1.0\linewidth,height=0.66\linewidth]{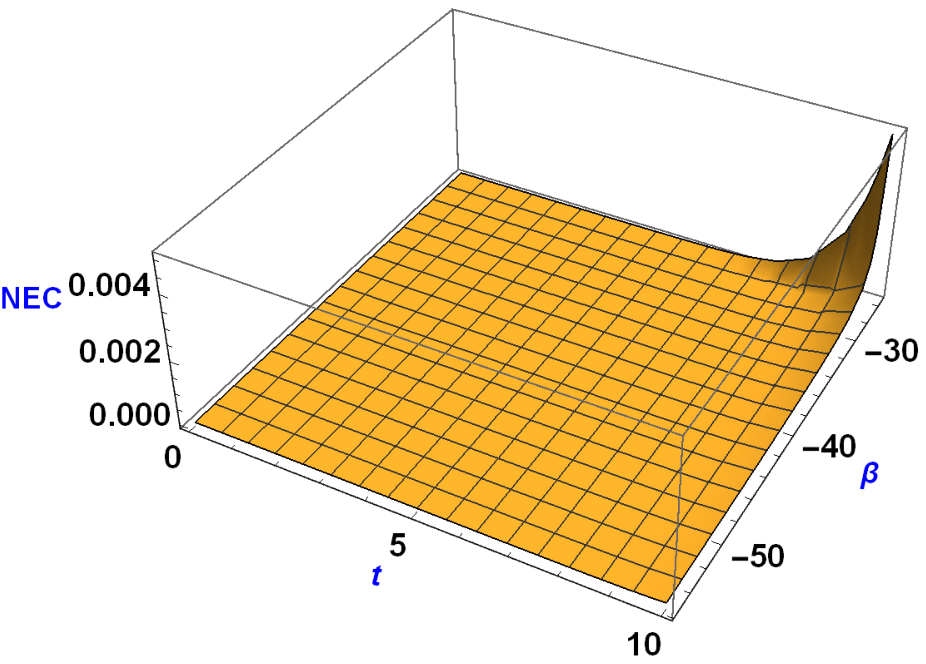}}\label{11b}
        \end{array}$
        \end{center}
        \begin{minipage}{\textwidth}
        \caption{Variations of NEC with $\a$, $\b$ and $t$.~$\text{Here $t$ is in Gyr.}$
        }\label{f11}
        \hrulefill
	\end{minipage}
        \begin{center}
        $\begin{array}{cc}
        \subfigure[] 
        {\includegraphics[width=1.0\linewidth,height=0.66\linewidth]{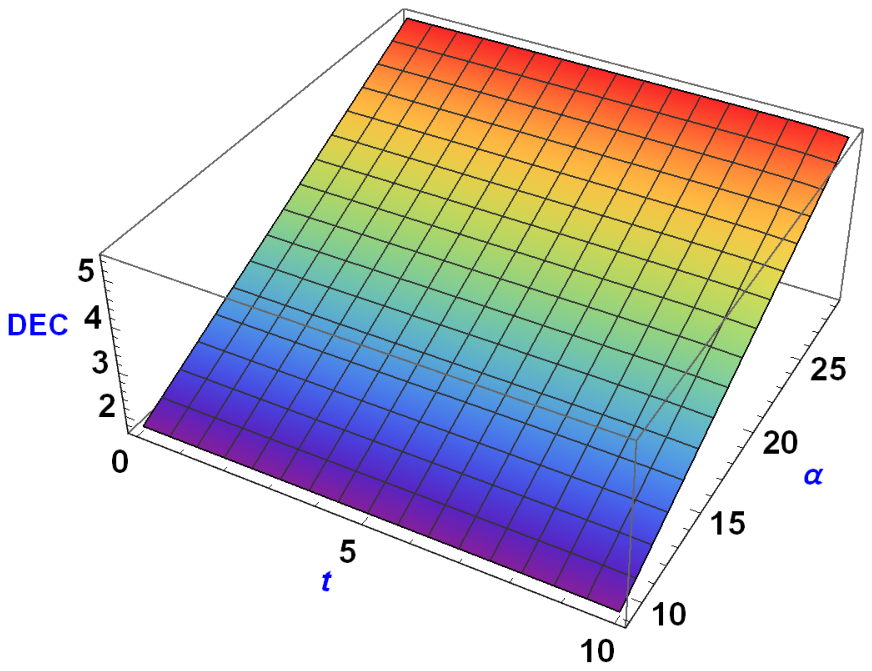}\label{12a}}
        \qquad
        \subfigure[] 
        {\includegraphics[width=1.0\linewidth,height=0.66\linewidth]{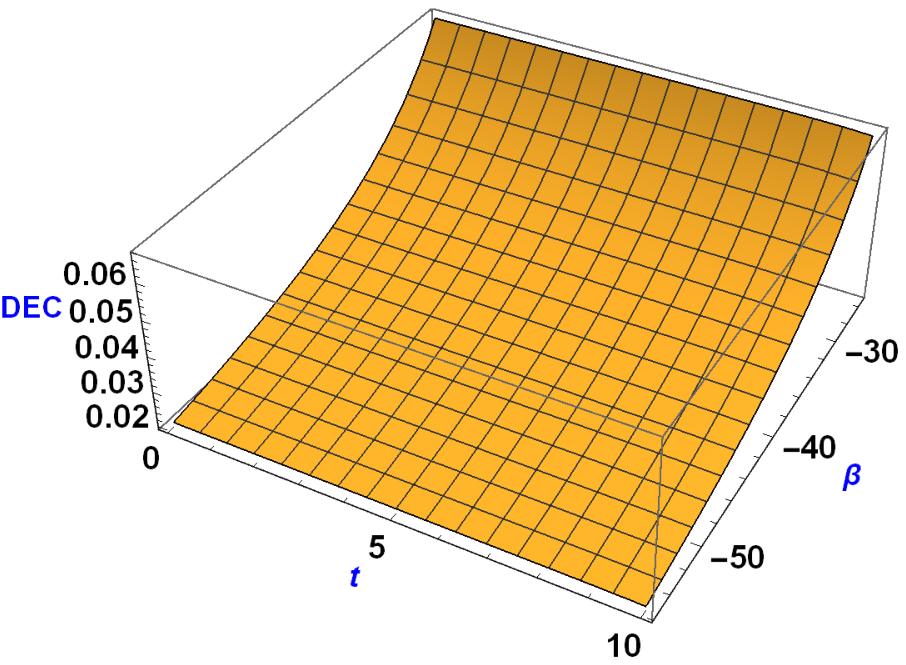}\label{12b}}
	\end{array}$
        \end{center}
        \begin{minipage}{\textwidth}
        \caption{Variations of DEC with $\a$, $\b$ and $t$.~$\text{Here $t$ is in Gyr.}$}\label{f12}
        \end{minipage}
    \end{figure}
\end{widetext}

    Among all the energy conditions, it is easy to see from Fig.\ref{f10} that the strong energy condition (SEC) is violating on the cosmological scale, which is obvious according to the recent data observations of the accelerating Universe \cite{Barcelo, Moraes1}. The variation in $\a$ and $\b$ results in the variation of SEC behaviour. Since the EOS parameter $\o$ is negative, it means $\r+3p<0$ and thus $\ddot{a}>0$, which implies there is a violation of SEC at present.

    On the other hand, the NEC and DEC both are obeying in this model (Figs.\ref{f11} and \ref{f12}). We have also observed the behaviour of NEC, which is a partial condition of WEC, as we have shown the behaviour of energy density in Fig.\ref{f4}. So, it is obvious that the validation of NEC and energy density together results in the validation of WEC.

\begin{figure}[H]
  \centering
  \includegraphics[width=8.5cm]{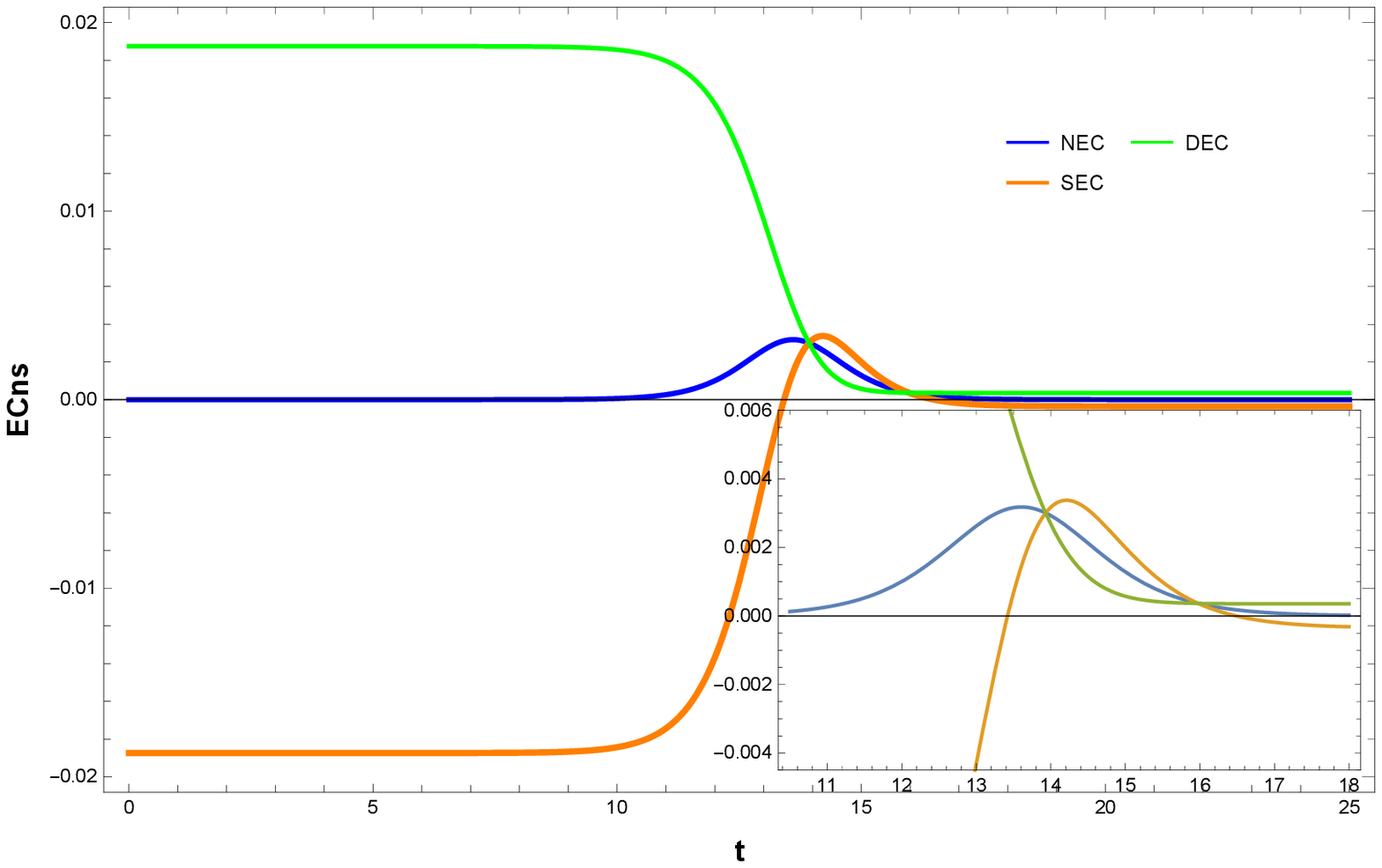}
  \caption{Plot of Energy Conditions as a function of cosmic time $(t\, \text{in Gyr})$ for $c=0.97$, $d=1$, $m=0.735$, $n=10$, $\a=0.1$, \& $\b=-59.1$}
  \label{f13}
\end{figure}

    The complete story of all the energy conditions is depicted together in Fig.\ref{f13}. From Fig.\ref{f13}, it is observed that the NEC and WEC satisfy the present model, but SEC violates. Since we've already discussed the profile of $\o$ previously, it starts with acceleration, then smoothly goes to the deceleration phase, and again it returns to the accelerated phase. The exciting thing is getting the same results for SEC, NEC, and DEC in Figs.\ref{f10}, \ref{f11}, and \ref{f12}, respectively.

\subsection{Case-B : $f(Q,T)=uQ^{\e+1}+v T$}
\noindent
    For the generic choice of $f(Q,T)=u Q^{\e+1} +v T$ gravity model (where $u,~v$ are arbitrary constants), we already have the expressions for the energy density $\r$ and pressure $p$ (Eq.\eqref{26}, and Eq.\eqref{27}). Like the previous case, similarly, we chose the model parameters $u=0.1,~v=-59.1$ with $\e=0.45$. Due to the importance of the energy conditions, again, we observed the acceptable ranges of model parameters $u$ are from $9.1$ to $29.1$, and for $v$, it is from $-25.55$ to $-55.55$.

\begin{widetext}
    \begin{figure}[H]
        \centering
	\begin{center} 
	$\begin{array}{cc}
	\subfigure[]
        {\includegraphics[width=1.0\linewidth,height=0.66\linewidth]{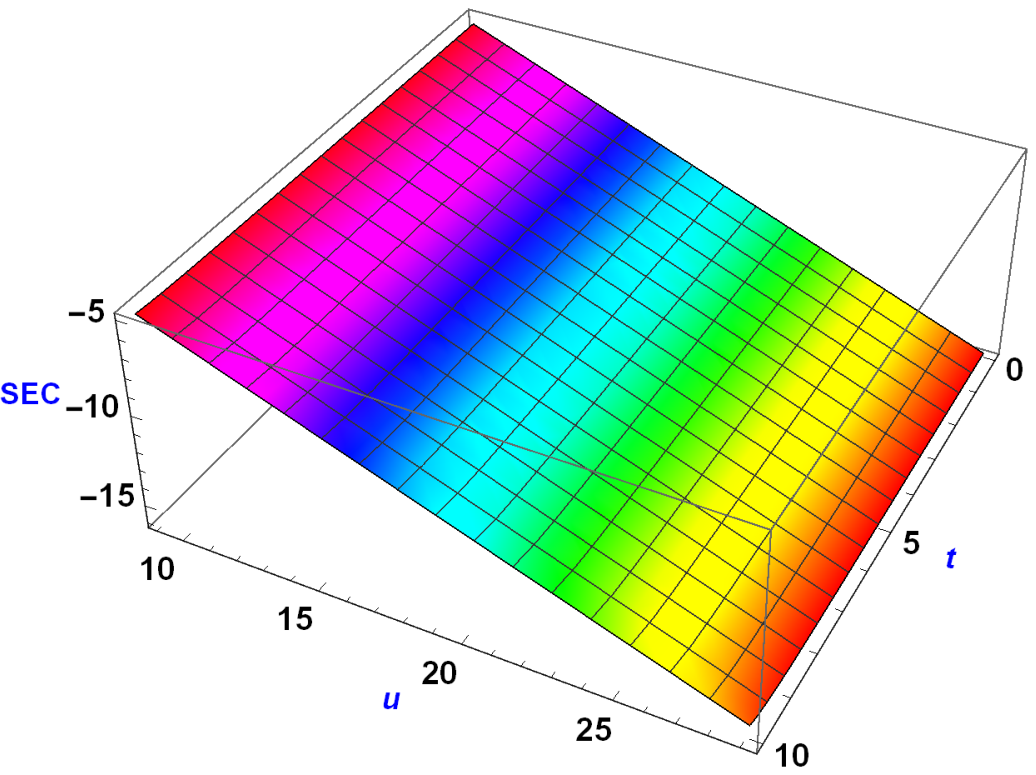}\label{14a}}
        \qquad
        \subfigure[]{\includegraphics[width=1.0\linewidth,height=0.66\linewidth]{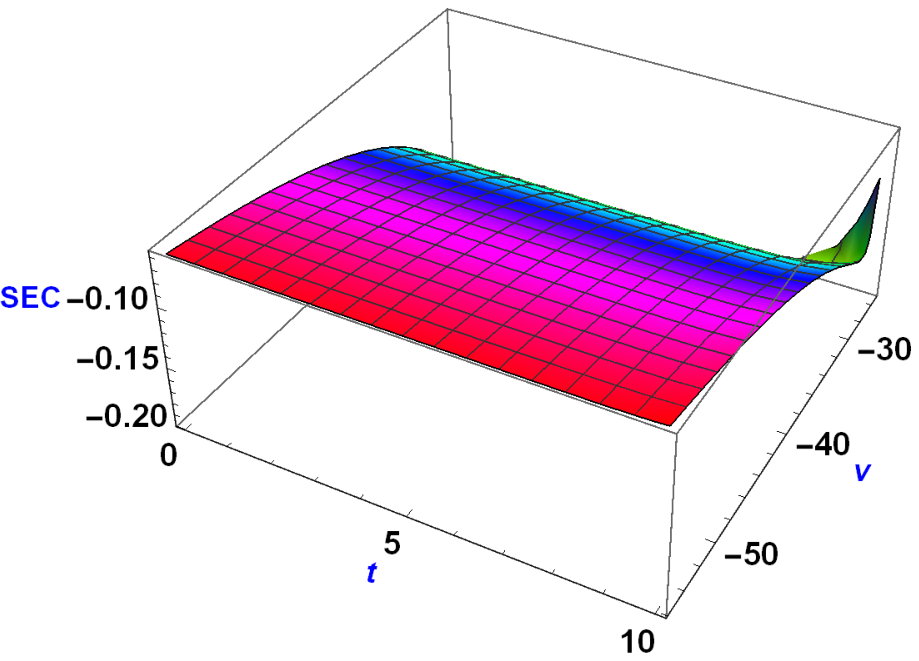}\label{14b}}
        \end{array}$
        \end{center}
        \begin{minipage}{\textwidth}
        \caption{Variation of SEC with model parameters $u,~v$ and cosmic time $t$.~$\text{Here $t$ is in Gyr.}$}\label{f14}
        \hrulefill
	\end{minipage}
        \begin{center}
        $\begin{array}{cc}
        \subfigure[] 
        {\includegraphics[width=1.0\linewidth,height=0.66\linewidth]{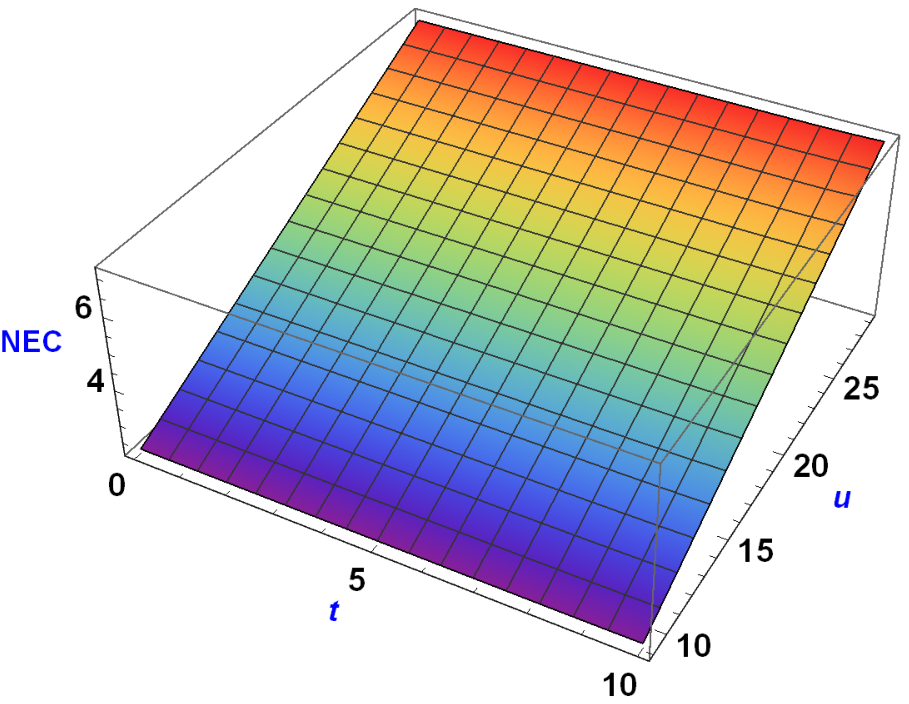}\label{15a}}
        \qquad
	\subfigure[] 
        {\includegraphics[width=1.0\linewidth,height=0.66\linewidth]{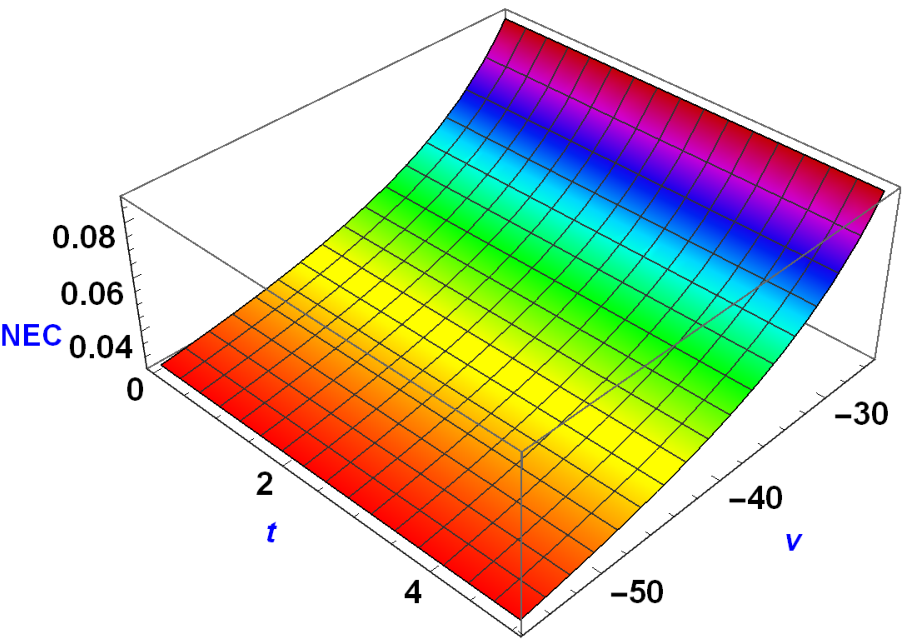}}\label{15b}
        \end{array}$
        \end{center}
        \begin{minipage}{\textwidth}
        \caption{Variation of NEC with model parameters $u,~v$ and cosmic time $t$.~$\text{Here $t$ is in Gyr.}$}\label{f15}
        \hrulefill
	\end{minipage}
        \begin{center}
        $\begin{array}{cc}
        \subfigure[] 
        {\includegraphics[width=1.0\linewidth,height=0.66\linewidth]{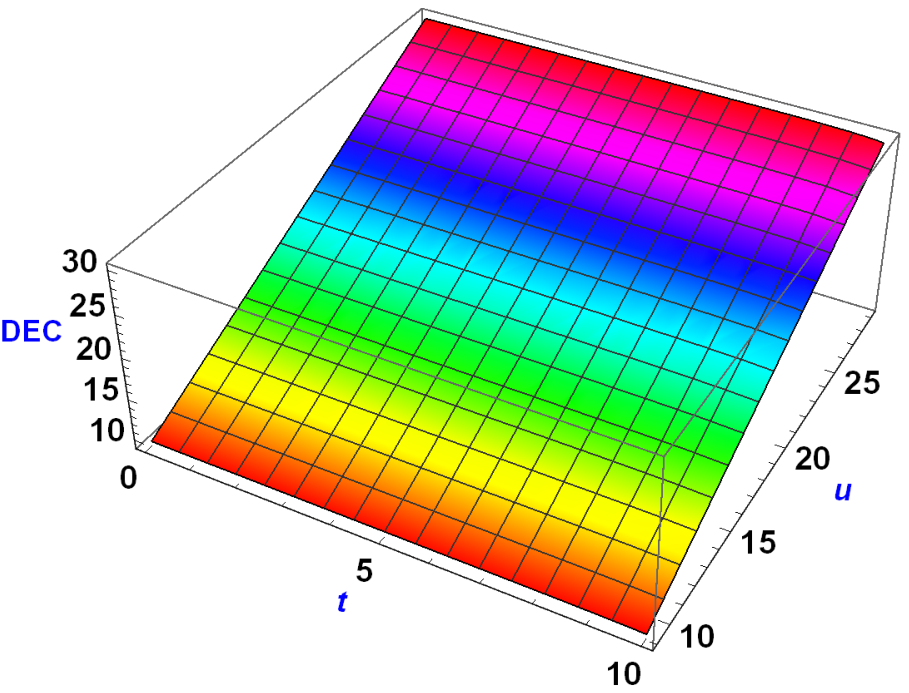}\label{16a}}
        \qquad
        \subfigure[] 
        {\includegraphics[width=1.0\linewidth,height=0.66\linewidth]{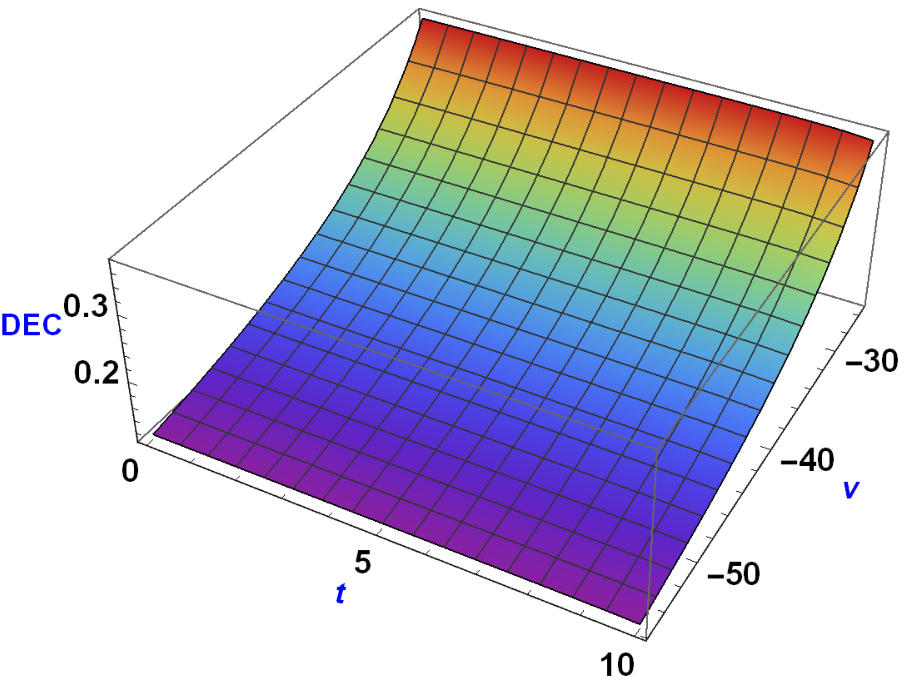}\label{16b}}
	\end{array}$
        \end{center}
        \begin{minipage}{\textwidth}
        \caption{Variation of DEC with model parameters $u,~v$ and cosmic time $t$.~$\text{Here $t$ is in Gyr.}$}\label{f16}
        \end{minipage}
    \end{figure}
\end{widetext}

    Again, among all the energy conditions, the strong energy condition is violated on the cosmological scale (Fig.\ref{f14}). The variation in $u$ and $v$ results in the variation of SEC behaviour. Since the EOS parameter $\o$ is negative, it means $\r+3p<0$, so there is a violation of SEC at the present epoch. The negative behaviour of SEC executes the accelerated expansion of the universe. 
    
    On the other hand, both the NEC and DEC do not violate this model. So the behaviour of NEC and DEC is always positive (Figs.\ref{f15} and \ref{f16}). Therefore, the validation of NEC and energy density together results in the validation of WEC.

\begin{figure}[H]
  \centering
  \includegraphics[width=8.5cm]{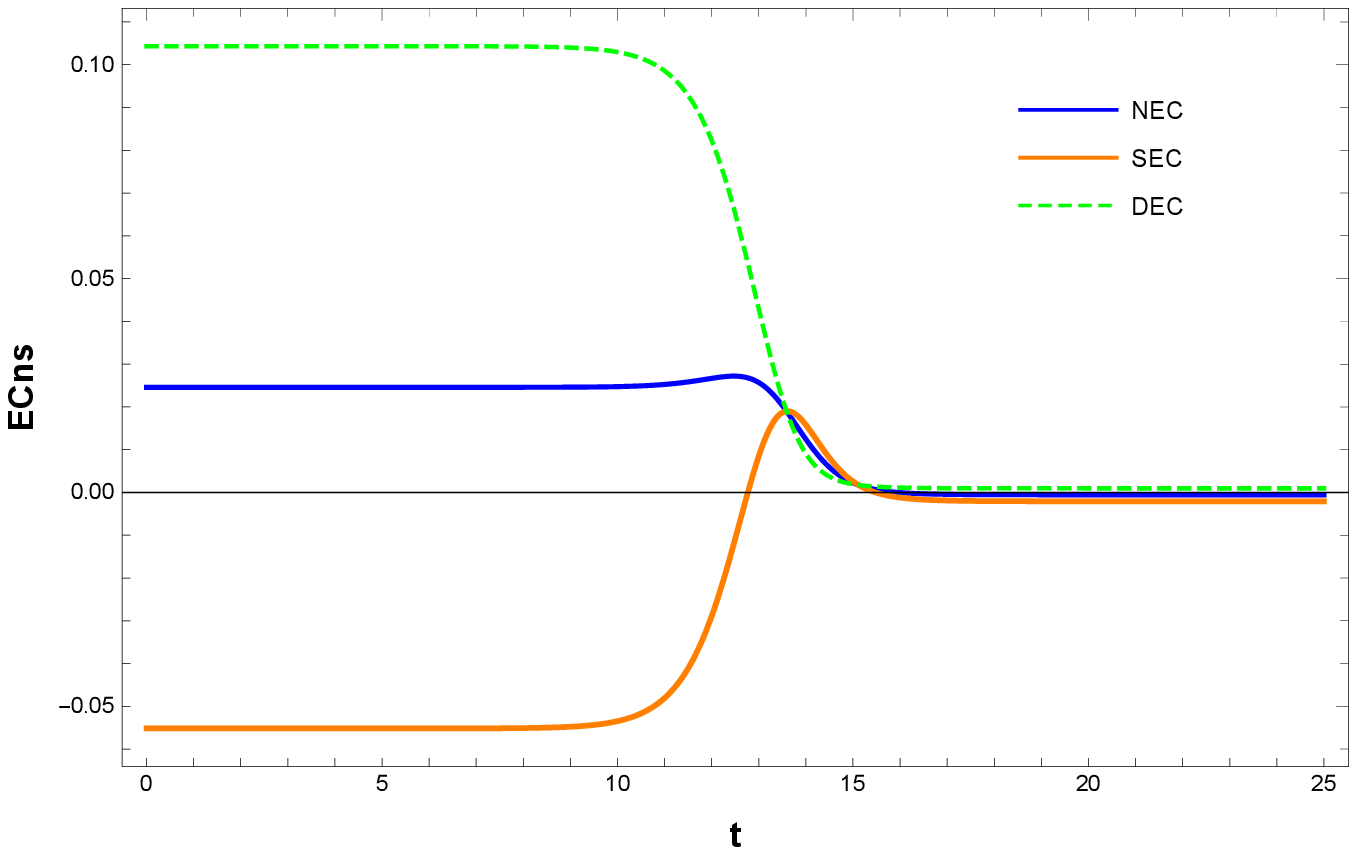}
  \caption{Plot of Energy Conditions as a function of cosmic time $t$ for $c=0.97$, $d=1$, $m=0.735$, $n=10$, $u=0.1$, \& $v=-59.1$}
  \label{f17}
\end{figure}

    Similarly, for the choice of $f(Q,T)=uQ^{1.45}+vT$ gravity model with the constants $u=0.1$ and $v=-59.1$, the portrait of all the energy conditions is shown in Fig.\ref{f17}. The same result is happening here, i.e., the violation of SEC with the non-violation of NEC and DEC, which satisfies the recent observation data of the accelerated expansion.

\section{Scalar Field Description}\label{VII}
\noindent
    In this section, we try to construct a scalar field description for the choice of scale factor Eq.\eqref{18} in $f(Q,T)$ gravity model for two different examples. This description is a more fundamental formulation for our study. To explain the early acceleration of the universe, i.e., the inflationary framework, the most appropriate and popular way is to consider a scalar field, which is known as inflation, managed by a specific potential. The action for the dynamics of inflation can be written as
\ben
    \mathcal{A}=\int d^4 x \sqrt{-g}\Big[\frac{R}{2\kappa}+L^{matt}_{\phi}\Big].
    \label{33}
\een
    Here $g$ is the determinant of the metric tensor $g_{\m\n}$, $R$ is the Ricci scalar, $\kappa=8\pi$ is a constant (setting $G=1$). The matter Lagrangian $L^{matt}_{\phi}$ contains an expression, which is minimally coupled of the inflaton field $\phi(t)$ to gravity, defined as follows.
\ben
    L^{matt}_{\phi}=-\frac{1}{2}g^{\m\n}\p_\m \phi\p_\n \phi -V(\phi),
    \label{34}
\een
    where $V(\phi)$ is the potential of the scalar field $\phi(t)$, that can depends on one or more free parameters \cite{Lyth}.

    The energy density $\r_{\phi}$ and the pressure $p_{\phi}$ in the scalar field model description can be described as
\ben
    \r_{\phi}=\frac{1}{2}{\dot{\phi}}^2(t)+V(\phi),~~~~~~~p_{\phi}=\frac{1}{2}{\dot{\phi}}^2(t)-V(\phi),\label{35}
\een
    where one can assume that the energy–momentum tensor of the inflaton field may behaves like a perfect fluid having a linear barotropic equation of state as $p_{\phi}=w\r_{\phi}$. On the other hand, the trace of the energy-momentum tensor in this scalar field description reads,
\ben
    T_{\phi}\equiv -\r_{\phi}+3p_{\phi}={\dot{\phi}}^2(t)-4V(\phi).\label{36}
\een
    
\subsection{Case-A}
\noindent
    For the linear model $f(Q,T)=\a Q+\b T$ using Eq.\eqref{35}, Eq.\eqref{15} and Eq.\eqref{16} can be rewritten as,
\ben
    &&3H^2=-\dfrac{(8\pi+\b){\dot{\phi_{1}^2}}(t)+2V(\phi_1)(8\pi+2\b)}{2\a},\label{37}
\\
    &&4\dot{H}+3H^2=\dfrac{3(8\pi+\b){\dot{\phi_{1}^2}}(t)-2V(\phi_1)(8\pi+2\b)}{2\a}.\non\\
    \label{38}
\een
    From Eq.\eqref{37}, and Eq.\eqref{38} one can simply find out the relation between the Hubble parameter and the inflaton field $\phi_1$ as,
\ben
    \dot{H}=\frac{{\dot{\phi}}^2_{1}}{2\a}(8\pi+\b),\label{39}
\een
    and we exclude the case for $\b=-8\pi$. Again, using Eq.\eqref{37}, and Eq.\eqref{39} we obtain the modified Klein-Gordon equation in this scalar field model as,
\ben
    \ddot{\phi_1}+3H\dot{\phi_1}+\frac{(8\pi+2\b)}{(8\pi+\b)}\frac{dV}{d\phi_1}=0.\label{40}
\een  

    Now by solving Eq.\eqref{39} with $\a=0.1$ and $\b=-59.1$, one may have the following:
\ben
    &&\phi_1(t)=-0.0767\tan^{-1}\big[\sinh(n-mt)\big],\label{41}
    \\
    &&H(\phi_1)=c-dm\sin(13.0378~\phi_{1}),\label{42}
    \\
    &&V(\phi_1)=0.0005\Big[6\big\{c-dm\sin(13.0378~\phi_1)\big\}^2\non\\
    &&-2~dm^2\cos^2(13.0378~\phi_1)\Big].\label{43}
\een

\begin{figure}[H]
  \centering
  \includegraphics[width=8.5 cm]{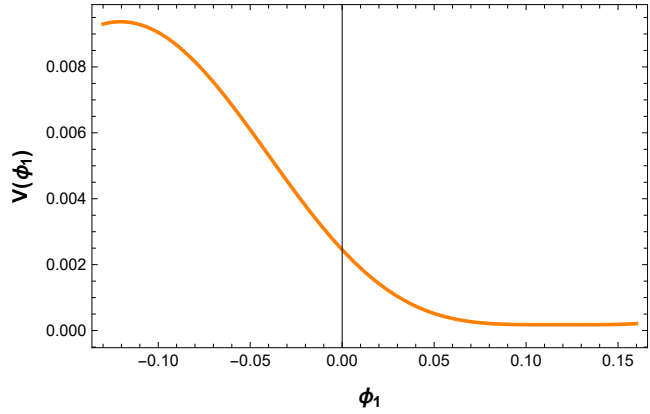}
  \caption{Plot of $V(\phi_1)$ with $\phi_1$ for $c=0.97$, $d=1$, $m=0.735$, \& $n=10$}
  \label{f18}
\end{figure}

    This potential $V(\phi_1)$ may be seen as a combination of the sine-cosine functions. The parametric plot of $V(\phi_1)$ vs. $\phi_1$ has been given in Fig.\ref{f18}. It is clear from Fig.\ref{f18} that during the early times, $V(\phi_1)$ is slowly varying with the inflaton field $\phi_1$ and at the end of the inflation, the field is slowly rolling down and it becomes constant in the dark energy dominated era i.e., in the late-times. We may conclude the behaviour of constant potential implies the cosmological constant-like behaviour during the late-times of the universe's evolution.
\vspace{0.5cm}

\subsection{Case-B}
\noindent
    Similarly, we obtain the effective energy density, effective pressure in more generic $f(Q,T)=u Q^{\e+1}+v T$ as follows:

\begin{widetext}
\ben
    &&3H^2\equiv 8\pi\r_{eff}=\dfrac{-(8\pi+v){\dot{\phi_{2}^2}}(t)-2V(\phi_2)(8\pi+2v)}{2u~(1+2\e)~(8\pi)^{\e}~\big({\dot{\phi_{2}^2}}+2V\big)^{\e}},\label{44}
\\
    &&2\dot{H}+3H^2\equiv -8\pi p_{eff}=\dfrac{(8\pi+v){\dot{\phi_{2}^2}}(t)-2V(\phi_2)(8\pi+2v)-\e\big((8\pi+v){\dot{\phi_{2}^2}}+2V(\phi_2)(8\pi+2v)\big)}{2u~(1+\e)(1+2\e)~(8\pi)^{\e}~\big({\dot{\phi_{2}^2}}+2V\big)^{\e}},\label{45}
\een    
\end{widetext}
\noindent
    and from Eq.\eqref{44}, and Eq.\eqref{45} one can simply obtain an important relation i.e.,
\ben
    H^{2\e}\dot{H}=\dfrac{(8\pi+v){\dot{\phi}}^2_{2}}{2u~6^{\e}(1+2\e)(1+\e)}.\label{46}
\een

    Now by considering a binomial expansion and neglecting the higher order terms in Eq.\eqref{46} and excluding $8\pi=-v$ case, we've solved Eq.\eqref{46} numerically with $u=0.1,~v=-59.1$ and $\e=0.45$. The expressions for the scalar field in terms of cosmic time $t$ is given by,
\ben
    \phi_{2}(t)=0.1874\Big[0.3421\text{sech}(n-mt)\non\\
    -\tan^{-1}\big[\sinh(n-mt)\big]\Big].\label{47}
\een
    
    Due to that extra term in the expression of $\phi_{2}(t)$, it is difficult to express the Hubble parameter in terms of $\phi_{2}$ only. But to proceed further, we introduced the scalar field $\phi_{1}(t)$ from the first model in our second model and did a comparison study between both of these models. So by using Eq.\eqref{41}, Eq.\eqref{47} can be written as,
\ben
    \phi_{2}(t)=\Big[0.0641\text{sech}(n-mt)+ 2.4432 \phi_{1}\Big].\label{48}
\een
    In this case, the Hubble parameter and the potential can also be written as,
\ben
    &&H=c+dm\sqrt{1-{\Big(\frac{\phi_{2}-2.4432\phi_{1}}{0.0641}\Big)}^2},\label{49}
    \\
    &&V=-\dfrac{u(1+2\e)6^{\e}H^{2\e}}{(2+\e)(8\pi+2v)}\Big[(1+\e)(2\dot{H}+3H^2)\non\\
    &&+3H^2+(1+\e)\e\dot{H}\Big].\label{50}
\een
    From the scalar field solution for our second model, one can observe that an additional term arises due to the change in the Lagrangian $f(Q,T)$ compared to the first model.  After some numerical analysis, it is observed that our model-1 is able to describe the inflationary cosmology better than model-2 due to its' linearity in the nature of Lagrangian. Further, we have seen that the scalar field solution for the first model dominates the second model because the scalar field description for the second model depends on the first model. In addition, we found that the potential is symmetric in nature for the second model, i.e., it is increasing initially and attends a maximum value when the inflation ends and later converges towards zero. In contrast, we may say that our linear model is the best compared to the non-linear model. 

\section{Cosmic Evolution in $\omega-\omega^{\prime}$ Phase Plane}\label{VIII}
\begin{figure}[H]
  \centering
  \includegraphics[width=8.5cm]{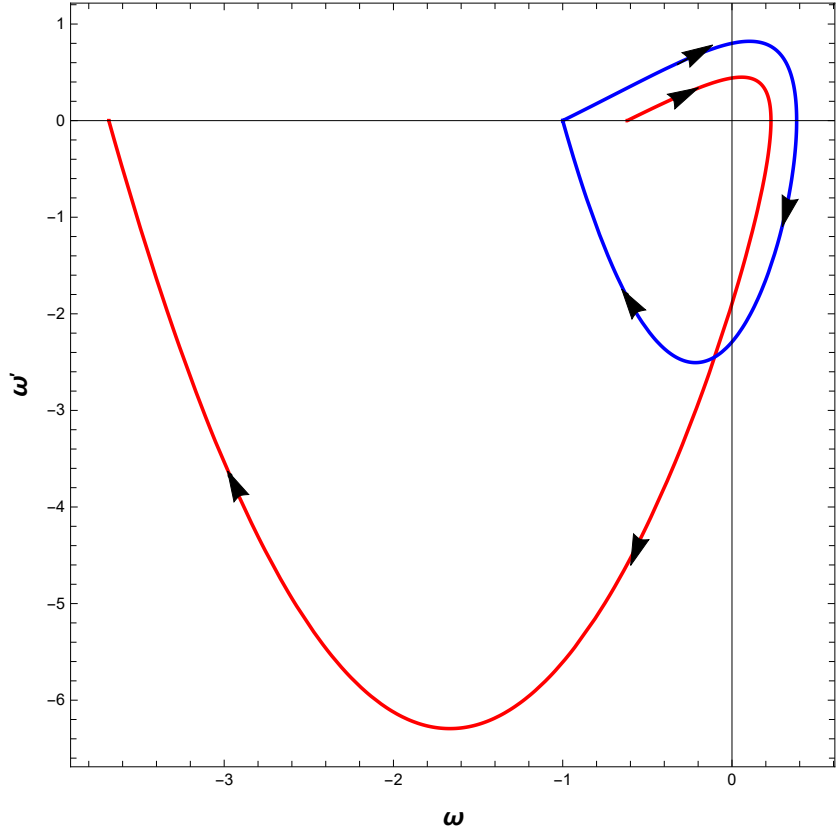}
  \caption{(color online) The evolutionary tracks in $(\o-\o^{\prime})$ phase space for $c=0.97$, $d=1$, $m=0.735$, $n=10$, $\a\equiv u=0.1$, \& $\b\equiv v=-59.1$}
  \label{f19}
\end{figure}

\noindent
    Cosmological evolution can also be presented in the well-known $\o-\o^{\prime}$ phase space representation. In this representation, by taking the scale factor as a parameter along the dynamics of the evolutionary path, one may define,
\ben
    \o^{\prime}=\frac{d\o}{d\ln a}.\label{51}
\een
    The evolutionary representation in this phase plane may be characterized into two regions, namely, the thawing region and the freezing region. The criterion for the identification of the thawing region is $\o^{\prime}>0$ and $\o<0$, and for the freezing region, it is given by $\o^{\prime}<0$ with $\o<0$. In the thawing region, $\o$ is an increasing function of time so that the evolution of the universe terminates into a de-Sitter-like stage. On the other hand, $\o$ decreases on time evolution in the freezing model. So its behaviour is asymptotic, and it will depend on the shape of the potential \cite{Caldwell,Linder,Sch}. Fig.\ref{f19} represents the evolution behaviour of the universe in $\o-\o^{\prime}$ phase space. In Fig.\ref{f19}, the blue curve represents our model one, i.e., $f(Q,T)=\a Q+\b T$, where the red trajectory defines the choice of the second model, $f(Q,T)=u Q^{\e+1}+v T$ with $\e=0.45$.

    From Fig.\ref{f19} (for the blue trajectory), it is clear that the universe starts with the acceleration smoothly in the thawing region and then produces its second phase of accelerated expansion in the freezing region during late times. In between these two regions, the EOS parameter shows some positive behaviour (goes with the maximum value of $\o\sim \frac{1}{3}$) in the decelerating era, where $\o^{\prime}$ will be first increasing and then decreasing. Similarly, the same story is happening for the red curve when we consider that the contribution of the power of $Q$ in the model is $1.45$. So, the freezing region is larger compared to the thawing region due to a higher non-metricity effect in the model. As a whole, one may say that during early times, the universe exhibits thawing behaviour, and now the universe is producing freezing behaviour.

\section{Statefinder Diagnostics}\label{IX}
\noindent
    As our universe is ever accelerating with unknown things, called dark energy, the unknown nature of dark energy raises many problems in modern cosmology. So, there always exists a question: what is dark energy? Many dark energy models such as $\Lambda$CDM, HDE, SCDM, CG, Quintessence, and K-essence have been proposed to understand the most intriguing nature of dark energy. These dark energy models have different behaviours in comparison to each other. The $\left\lbrace r,s\right\rbrace$ parametrization technique is used to distinguish all these kinds of dark energy models \cite{Sahni,Alam}.\\
    One can define $r$ and $s$ as follows:
\ben
    &&r=\frac{\dot{\ddot{a}}}{aH^3},\label{52}
    \\
    &&s=\frac{2}{3}\frac{r-1}{2q-1},\label{53}
\een
    where $q\neq \frac{1}{2}$. Now, the different pairs of $\left\lbrace r,s\right\rbrace$ represent different kinds of dark energy models.

\begin{itemize}
    \item[(a).] For the $\Lambda$CDM model, the pair is $\left\lbrace r=1,s=0\right\rbrace$.

    \item[(b).] For the HDE model, the pair is $\left\lbrace r=1,s=\frac{2}{3}\right\rbrace$..

    \item[(c).] For the SCDM model, the pair is $\left\lbrace r=1,s=1\right\rbrace$. 

    \item[(d).] For the Quintessence model, the pair is $\left\lbrace r<1,s>0\right\rbrace$.

    \item[(e).] For the CG model, the pair is $\left\lbrace r>1,s<0\right\rbrace$. 
\end{itemize}

    We need to study the convergence and divergence nature of the trajectory of the $r-s$ curve corresponding to any cosmological dark energy models. So clearly, the deviation from $\left\lbrace 1,0 \right\rbrace$ represents the deviation from $\Lambda$CDM model. Furthermore, from the observation \cite{Albert,Albert2}, the values of $r$ and $s$ could be concluded. Obviously, it is worthy to describe the various dark energy models in the near future.

\begin{figure}[H]
  \centering
  \includegraphics[width=8.5 cm]{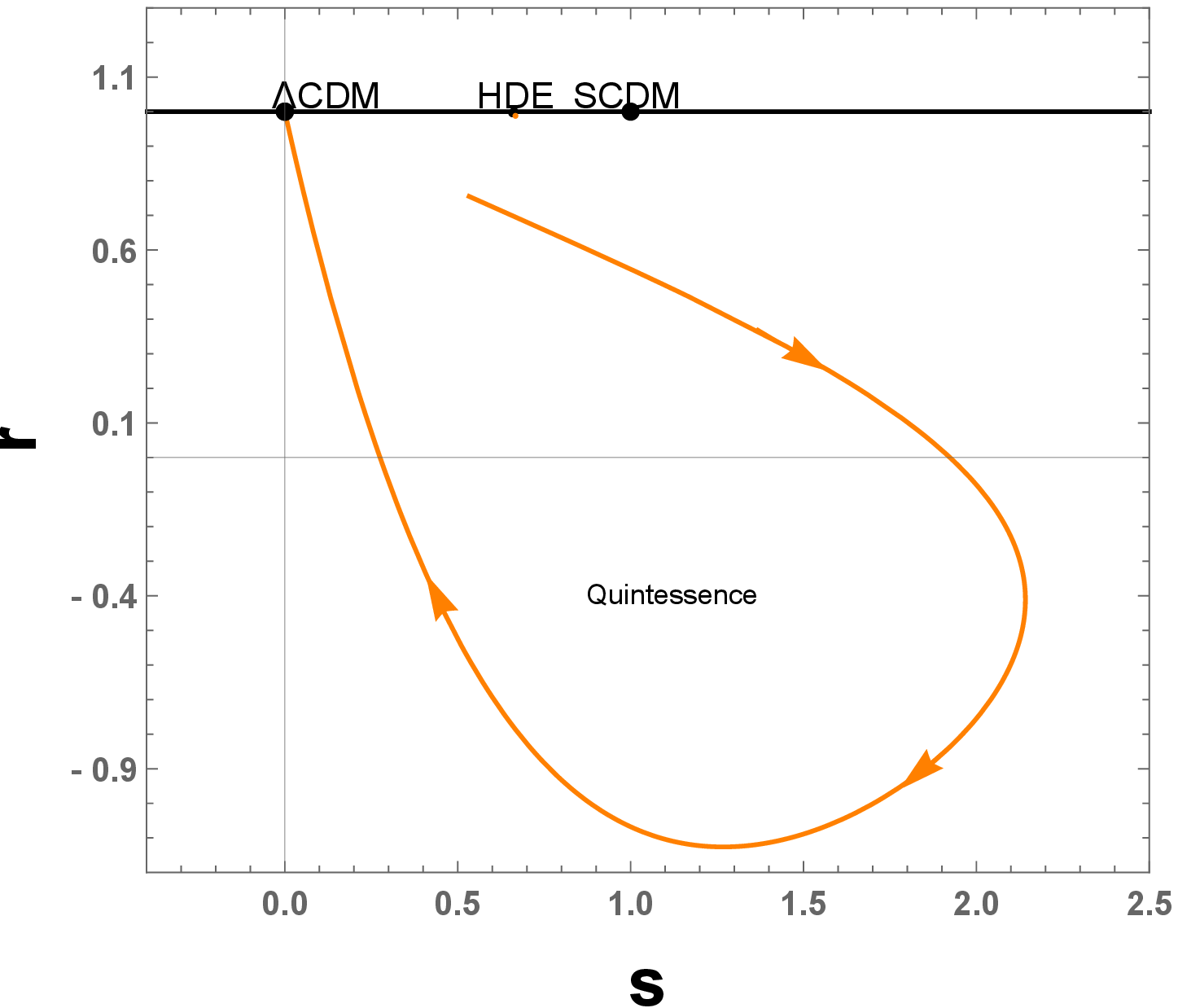}
  \caption{$r-s$ parametric plot for $c=0.97,d=1,m=0.735,$ \& $n=10$.}
  \label{f20}
\end{figure}

\begin{figure}[H]
  \centering
  \includegraphics[width=8.5 cm, height=12.0 cm]{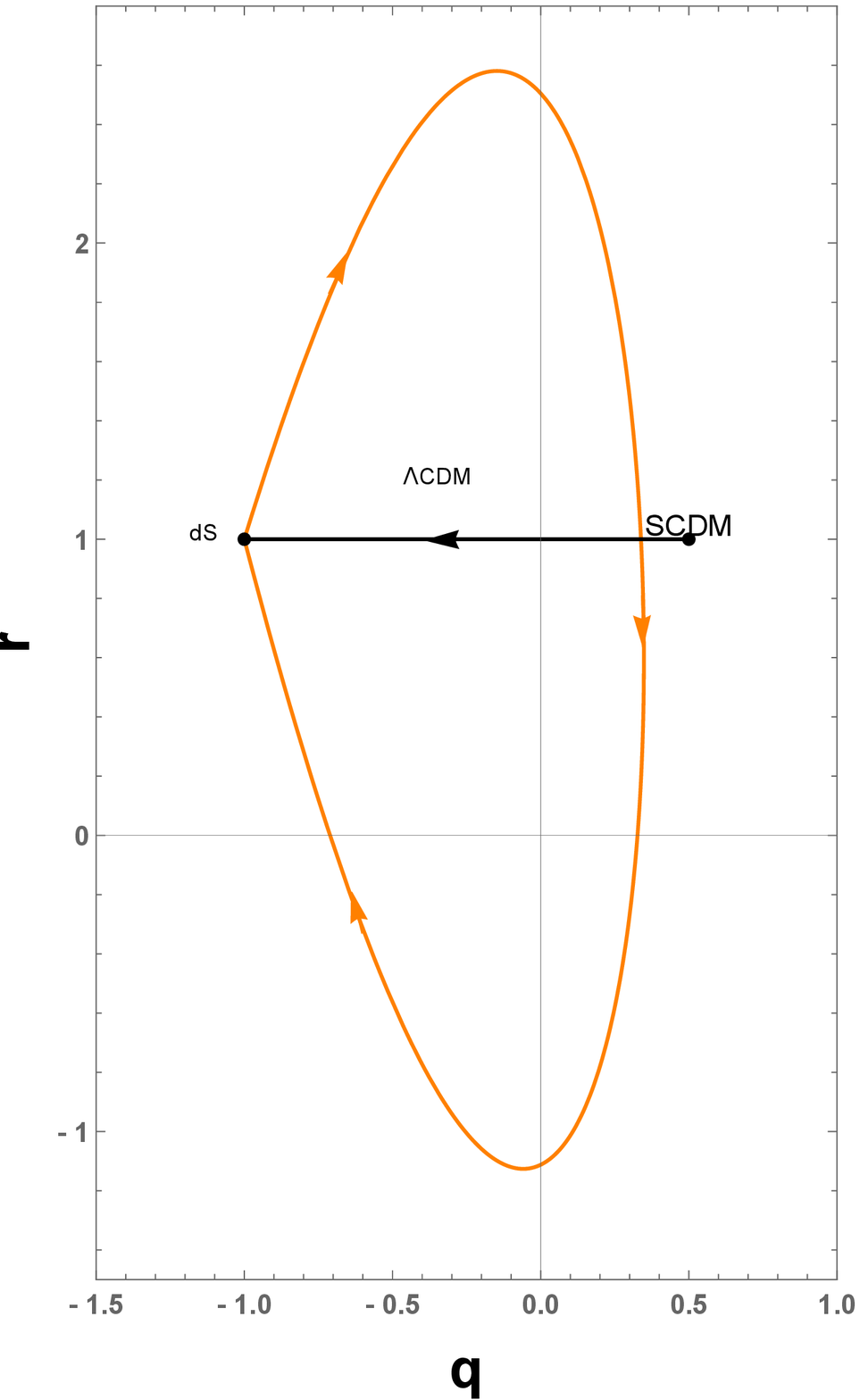}
  \caption{$r-q$ parametric plot for $c=0.97,d=1,m=0.735,$ \& $n=10$.}
  \label{f21}
\end{figure}

    Now using Eq.\eqref{18} in Eq.\eqref{52}, and Eq.\eqref{53}, we can rewrite the $r,s$ as given below (Eq.\eqref{54} and Eq.\eqref{55}). In Fig.\ref{f20}, the parametrization of $r$ and $s$ are shown in the $(r,s)$ plane, and the arrow mark represents the direction of the trajectory. From Fig.\ref{f20}, it is clearly observed that, initially, the trajectory diverges from the $\Lambda$CDM model, and later, it converges to the $\Lambda$CDM model. Also, the evolution of the trajectory completely lies in the quintessence era. On the other hand, we have shown the parametrization of $r$ and $q$ in Fig.\ref{f21}. From Fig.\ref{f21}, we observed that our model starts with the de-Sitter universe, and initially, it goes to Chaplygin gas, represented by $r>1$, and after that, it comes to quintessence and finally again converges to the de-Sitter universe.

\begin{widetext}
\ben
    &&r=\frac{c^3+d m \tanh (n-m t) \left\lbrace c^2+(d+1) m \tanh (n-m t) [3 c+(d+2) m \tanh (n-m t)]-(3 d+2) m^2\right\rbrace-3 c d m^2}{[c+d m \tanh (n-m t)]^3},\non
    \\
    \label{54}
    \\
    &&s=\frac{2 d m^2 \text{sech}^2(n-m t) [3 c+(3 d+2) m \tanh (n-m t)]}{3 [c+d m \tanh (n-m t)] \left\lbrace c^2+d m \tanh (n-m t) [6 c+(3 d+2) m \tanh (n-m t)]-2 d m^2\right\rbrace}.\non
    \\
    \label{55}
\een
\end{widetext}

\section{Conclusion}\label{X}
\noindent
    In this manuscript, we embark on an exploration of a novel cosmological framework within the framework of $f(Q,T)$ gravity. The universe has traversed various evolutionary phases from its early inflation to the present day. Numerous cosmological models have been proposed to analyze these phases within the realm of gravitational theories. Nevertheless, there exists an ongoing quest for a comprehensive cosmological model capable of elucidating 

\noindent
    the entire evolution of the universe, commencing from the early inflationary epoch and culminating in the late-time cosmic acceleration. To address this challenge, we introduce two distinct $f(Q,T)$ cosmological models and conduct a detailed examination of their matter evolution profiles along with the scrutiny of various energy conditions.

    The equation of state parameter $(\omega)$ is pivotal in characterizing the diverse properties of cosmic fluids within a cosmological model. Distinct values of $\omega$ correspond to varying matter profiles, encompassing scenarios like radiation-dominated $(\omega=1/3)$, matter-dominated $(\omega=0)$, quintessence $(1/3<\omega<-1)$, $\Lambda$CDM $(\omega=-1)$, and phantom $(\omega<-1)$. Consequently, we undertake a thorough analysis of the equation of state parameters across cosmic epochs for both of the cosmological models.

    Our observations reveal that both models exhibit accelerated expansion during early and late-time cosmic evolution. Specifically, Model-I demonstrates a $\Lambda$CDM-like expansion during early and late epochs, whereas Model-II exhibits quintessence-like behaviour in the early universe and evolves toward a phantom-like expansion in late times. This late-time accelerated expansion can be attributed to a substantial amount of negative pressure within the universe. The comprehensive evolution of the equation of state parameter delineates a trajectory commencing with early-time inflation, transitioning into a decelerated phase, and subsequently entering a second phase of accelerated cosmic expansion. Further, one can observe some bouncing profiles because we are exploring early and late-time expansion of the universe through a single cosmological model. When the model shifts from an inflationary-dominated phase to late-time acceleration through the matter-dominated phase, at that moment we can observe the bouncing behaviour. We can see this type of evolution in pressure ($p$) and equation of state parameters ($\omega$) as these profiles play an important role in the evolution process, whereas density profiles evolve smoothly as time goes on for both models.
    
    Furthermore, we subject our cosmological models of energy conditions to assess their viability. Across all energy condition profiles, we consistently observe positive energy densities throughout the cosmic evolution. The Weak Energy Condition (WEC), Null Energy Condition (NEC), and Dominant Energy Condition (DEC) are satisfied by both models. However, the Strong Energy Condition (SEC) is violated during the early and late-time evolution, which aligns with the theory of accelerated expansion. Intriguingly, both models conform to the SEC during the decelerated phase of cosmic evolution. Additionally, we re-evaluate the behaviour of dark energy in our cosmological solutions through a $\omega-\omega'$ phase space, characterized by thawing and freezing regions, and employ statefinder diagnostics. These findings align with a scalar field description for the models within the framework of $f(Q,T)$ gravity. We also discern that the potential $V(\phi_1)$ maintains a constant value as $\phi_1$ varies, indicative of cosmological-constant-like behaviour in the late stages of cosmic expansion. Notably, the first model exhibits a more compelling explanation for inflationary epochs compared to the second model, primarily due to the presence of high non-linear effects in the Lagrangian of the latter. Further, we would like to note here that we have chosen the model parameters values as per the previous studies in $f(Q,T)$ with the observational agreement to present the accelerated expansion of the universe, whereas the values of free parameters of scale factor are chosen in to present the desirable scenario of the cosmological model.

    The idea investigated in our study is the reason to advance over previous studies in the context of $f(Q,T)$ gravitational theories. For instance, the studies have been done in this setup by focusing on a particular cosmological scenario, either to study the late-time acceleration \cite{Xu,Simran3,Simran2,gaurav} or the early-time cosmological scenario \cite{Maryam}. We have presented a complete evolution profile of the universe with all important energy-dominated phases in the background of $f(Q,T)$ gravity. Moreover, it is important to mention here that the recent studies on parametrization on various cosmological parameters are interesting as a future perspective that can explore a time-dependent dark energy model in the context of diverse modified theories of gravity \cite{G8,G9,G10,G11,G12,G13,G14,G15}.

    In conclusion, our models offer a comprehensive portrayal of the universe's evolution, spanning from its early inception to late-time cosmic epochs, within the context of $f(Q,T)$ gravity. We anticipate that this study will contribute valuable insights into the prospect of elucidating the entire evolutionary trajectory of the universe through a cosmological model. In the future, it would be intriguing to subject these models to observational tests, which may lead to the discovery of new physics and a deeper understanding of the cosmos. For instance, confronting a cosmological model against the observational datasets indicating early expansion and late-time expansion and comparing them may bring interesting results.
\\ \\
\textbf{Acknowledgement}
    We would like to express our sincere gratitude to Sergey Odintsov, Kazuharu Bamba, Shin’ichi Nojiri, Vasilis Oikonomou, and Raj Kumar Das for their invaluable discussions and insightful comments on this manuscript. SM gratefully acknowledges the Japan Society for the Promotion of Science (JSPS) for providing the postdoctoral fellowship for 2024-2026 (JSPS ID No.: P24026). This work of SM is supported by the JSPS KAKENHI Grant (Number: 24KF0100). We are also deeply appreciative of the constructive feedback provided by the referee and the editor, which significantly enhanced the quality of our work.
\\ \\
\textbf{Author contributions}
    All authors have accepted responsibility for the entire content of this manuscript and approved its submission.
\\ \\
\textbf{Data availability} 
    Data sharing is not applicable to this article as no datasets were generated or analysed during the current study.
\\ \\
\textbf{Conflict of interest} 
    Authors state no conflicts of interest.
\\ \\
\textbf{Informed consent} 
    Informed consent was obtained from all individuals included in this study.

\end{document}